%% file: pmhmc_arxiv2.tex
\documentclass[english,11pt]{article}
\usepackage[lmargin=1in,rmargin=1in,tmargin=0.95in]{geometry}

\usepackage{abbreviations}

\usepackage[T1]{fontenc}
\usepackage{lmodern}
\usepackage{hyperref}       % hyperlinks
\usepackage{url}            % simple URL typesetting
\usepackage{booktabs}       % professional-quality tables
\usepackage{amsfonts}       % blackboard math symbols
\usepackage{nicefrac}       % compact symbols for 1/2, etc.
\usepackage{microtype}      % microtypography

\usepackage[compress]{natbib}
\usepackage{graphicx}
\usepackage{enumitem}
\usepackage{algorithm}
\usepackage{float}
\usepackage{amsmath}
\usepackage{amsthm}
\usepackage{amssymb}
\usepackage{xspace}
\usepackage{mathtools}

\let\originalleft\left
\let\originalright\right
\renewcommand{\left}{\mathopen{}\mathclose\bgroup\originalleft}
\renewcommand{\right}{\aftergroup\egroup\originalright}

\graphicspath{{./}{./figs/}{./figs/glmm/}{./figs/sinc2traces/}}

\usepackage{color}

\theoremstyle{plain}
\newtheorem{proposition}{Proposition}
\theoremstyle{plain}

\theoremstyle{plain}
\newtheorem{lemma}{Lemma}
\theoremstyle{plain}

\theoremstyle{remark}
\newtheorem*{remark}{Remark}

\newcommand{\ja}[1]{}
\newcommand{\fl}[1]{}

\input{defs_jmlr}

\title{Pseudo-Marginal Hamiltonian Monte Carlo}
\date{}
\author{Johan Alenl\"ov\thanks{Department of Information Technology, Uppsala University, Sweden.}, Arnaud Doucet\thanks{Department of Statistics, Oxford University, UK.}, and  Fredrik Lindsten\thanks{Division of Statistics and Machine Learning, Link\"oping University,	Sweden.}}

\begin{document}

\maketitle
\begin{abstract}%
Bayesian inference in the presence of an intractable likelihood
function is computationally challenging. When following a Markov chain
Monte Carlo (MCMC) approach to approximate the posterior distribution in
this context, one typically either uses MCMC schemes which target
the joint posterior of the parameters and some auxiliary
latent variables, or pseudo-marginal Metropolis--Hastings (MH) schemes.
The latter mimic a MH algorithm targeting the marginal posterior of the
parameters by approximating unbiasedly the intractable likelihood. However, in scenarios where the parameters and auxiliary variables
are strongly correlated under the posterior and/or this posterior
is multimodal, Gibbs sampling or Hamiltonian Monte Carlo (HMC) will
perform poorly and the pseudo-marginal MH algorithm, as any other MH scheme, will be
inefficient for high dimensional parameters.
We propose here an original MCMC algorithm, termed pseudo-marginal
HMC, which combines the advantages of both HMC and pseudo-marginal schemes. 
Specifically, the pseudo-marginal HMC method is controlled by a precision parameter $N$, controlling 
the approximation of the likelihood and, for any $N$, it samples the \emph{marginal
posterior} of the parameters. Additionally, as $N$ tends to infinity, its
sample trajectories and acceptance probability converge to those of an ideal, but intractable, HMC algorithm which 
would have access to the marginal posterior of parameters and its gradient.
We demonstrate through experiments that
pseudo-marginal HMC can outperform significantly both standard HMC
and pseudo-marginal MH schemes.
\end{abstract}

%\begin{keywords}
%	Hamiltonian Monte Carlo, pseudo-marginal, Markov chain Monte Carlo, latent variable models.
%\end{keywords}

\section{Introduction\label{sec:Introduction}}

Let $y\in\mathcal{Y}$ denote some observed data and $\theta\in\Theta\subseteq\mathbb{R}^{d}$
denote parameters of interest. We write $\theta\mapsto p(y\mid\theta)$ for
the likelihood of the observations and we assign a prior for $\theta$
of density $p\left(\theta\right)$ with respect to Lebesgue measure
$\mathrm{d}\theta$. Hence the posterior density of interest is given
by
\begin{equation}
\pi(\theta)=p\left(\theta\mid y\right)\propto p(y\mid\theta)p\left(\theta\right).\label{eq:posterior}
\end{equation}
For complex Bayesian models, the posterior (\ref{eq:posterior})
needs to be approximated numerically. When the likelihood $p(y\mid\theta)$
can be evaluated pointwise, this can be achieved using standard MCMC
schemes. However, we will consider here the scenario where $p(y\mid\theta)$
is intractable, in the sense that it cannot be evaluated pointwise.
We detail below two important scenarios where an intractable likelihood
occurs.

\vspace{1ex}
\noindent
\textbf{Example:\ Latent variable models. }Consider observations $y=y_{1:T}$
such that
\begin{align}
X_{k}&\overset{\text{i.i.d.}}{\sim}f_{\theta}(\cdot)\text{, }
&
\left.Y_{k}\right\vert X_{k}&\sim g_{\theta}(\cdot\mid X_{k}), \label{eq:independentlatentvariablemodels}
\end{align}
where $\left(X_{k}\right)_{k\geq1}$ are $\mathbb{R}^{n}$-valued
latent variables, $\left(Y_{k}\right)_{k\geq1}$ are $\mathsf{Y}$-valued (thus, $\mathcal{Y}= \mathsf{Y}^T$)
and $i:j:=\left\{ i,i+1,...,j\right\} $ for any $i<j$.
Having observed $Y_{1:T}=y_{1:T}$, the likelihood is given by
\(
p(y_{1:T}\mid\theta)= \prod_{k=1}^{T} p(y_{k}\mid\theta),
\)
where each term $p(y_{k}\mid\theta)$ satisfies
\begin{equation}
p(y_{k}\mid\theta)=\int_{\mathbb{R}^{n}}\,f_{\theta}(x_{k}) g_{\theta}(y_{k}\mid x_{k})\mathrm{d}x_{k}.\label{eq:integrallikelihoodpaneldata}
\end{equation}
If the integral (\ref{eq:integrallikelihoodpaneldata}) cannot be
computed in closed form then the likelihood $p(y_{1:T}\mid\theta)$
is intractable.

\vspace{1ex}
\noindent
\textbf{Example:\ Approximate Bayesian computation (ABC).} Consider the scenario where $\theta\mapsto\widetilde{p}(y\mid\theta)$
is the ``true'' \ likelihood function. We cannot
compute it pointwise but we assume we are able to simulate some pseudo-observations
$Z\sim\widetilde{p}(\cdot\mid\theta)$ using $Z=\gamma\left(\theta,V\right)$
where $V\sim\lambda\left(\cdot\right)$
for some auxiliary variable distribution $\lambda$ and mapping
$\gamma:\Theta\times\mathcal{V\rightarrow Y}$.
Given a kernel $K:\mathcal{Y\times Y\rightarrow}\mathbb{R}^{+}$,
the ABC approximation of the posterior is given by (\ref{eq:posterior})
where the intractable ABC\ likelihood is
\begin{align}
p(y  \mid\theta)=\int_{\mathcal{Y}}K(\left.y\right\vert z)\widetilde{p}(z\mid\theta)\mathrm{d}z
=\int K(\left.y\right\vert \gamma\left(\theta,v\right))\lambda\left(v\right)\mathrm{d}v.
\label{eq:likelihoodABC}
\end{align}
\vspace{1ex}

Two standard approaches to perform MCMC\ in these scenarios are:
\begin{enumerate}[leftmargin=1em]
\item Implement standard MCMC\ algorithms to sample from the joint distribution
of the parameters and auxiliary variables; e.g. in the latent variable
context we would target $p\left(\left.\theta,x_{1:T}\right\vert y_{1:T}\right)\propto p\left(\theta\right){\textstyle \prod\nolimits _{k=1}^{T}}\,f_{\theta}(x_{k})g_{\theta}(y_{k}\mid x_{k})$
and in the ABC\ context $p(\theta,v\mid y)\propto p\left(\theta\right)\lambda\left(v\right)K\left(\left.y\right\vert \gamma\left(\theta,v\right)\right)$.
Gibbs type approaches sampling alternately the parameters and the
auxiliary variables can converge very slowly if these variables are
strongly correlated under the target \citep[Section 2.3]{andrieudoucetholenstein2010}.
Hamiltonian Monte Carlo (HMC) methods \citep{duane1987} offer a possible remedy, but can also struggle in cases where there are strong dependencies between variables, or when the joint posterior is multimodal \citep[Section 5.5.7]{Neal2011}.
\item Use a pseudo-marginal MH algorithm which replace the intractable likelihood
term by a non-negative unbiased estimate of the true likelihood; see
\citep{andrieudoucetholenstein2010,andrieu2009pseudo,beaumont2003estimation,fluryshephard2011,linliuSloan2000}.
For example, in the ABC context the pseudo-marginal MH algorithm is
a MH algorithm targeting $p(\theta,v\mid y)$ using a proposal distribution
$q\left(\theta,\theta^{\prime}\right)\lambda\left(v^{\prime}\right)$.
As for any MH\ algorithm, it can be difficult to select a proposal
$q\left(\theta,\theta^{\prime}\right)$ which results in an efficient sampler when $\Theta$ is high-dimensional.
\end{enumerate}
In many scenarios, the marginal posterior (\ref{eq:posterior}) will have a ``nicer'' structure than the joint posterior of the parameters
and auxiliary variables which often exhibit complex patterns of dependence and multimodality. For example, discrete
choice models are a widely popular class of models in health economics,
e-commerce, marketing and social sciences used to analyze choices
made by consumers/individuals/businesses \citep{Train2009}. When
the population is heterogeneous, such models can be represented as
(\ref{eq:independentlatentvariablemodels}) %-(\ref{eq:likelihoodpaneldata})
where $f_{\theta}(\cdot)$ is a mixture distribution, the number of
components representing the number of latent classes; see e.g. \citep{burda2008}.
In this context, the paucity of data typically available for each
individual is such that the joint posterior $p(\theta,x_{1:T}\mid y_{1:T})=p(\theta\mid y_{1:T}){\textstyle \prod\nolimits _{k=1}^{T}}p(x_{k}\mid y_{k},\theta)$
will be highly multimodal while the marginal $p(\theta\mid y_{1:T})$ will
only have symmetric well-separated modes for $T$ large enough, or one mode if constraints on the mixture parameters are introduced. Such problems also
arise in biostatistics \citep{komarek2008}. In these scenarios, current
MCMC methods will be inefficient. In this article, we propose a novel
HMC\ scheme, termed pseudo-marginal HMC, which mimics the HMC\ algorithm
targeting the marginal posterior (\ref{eq:posterior}) while integrating out
numerically the auxiliary variables.
The method is a so called \emph{exact approximation} in the sense that its limiting distribution (marginally in $\theta$) is precisely $\pi(\theta)$.

\subsection{Related work}
Stochastic gradient MCMC~\citep{Welling:2011,chen2014,Ding:2014,leimkuhler2016}---including HMC-like methods---are a popular class of algorithms for approximate posterior sampling when an unbiased estimate of the \emph{log-likelihood gradient} is available.
The typical scenario is when the number of data points $T$ is prohibitively large for evaluating the full gradient, in which case sub-sampling (mini-batching) can be used to approximate the gradient unbiasedly. This is in contrast with the setting studied in this paper, where we assume that we have access to an unbiased estimate of the likelihood itself, but not of the log-likelihood gradient.
Furthermore, these methods are inconsistent for finite step sizes and typically require some type of variance reduction techniques to be efficient \citep{Shang:2015}. Recently, \citet{umenberger2019} have proposed to use debiasing \citep{jacob:2019} of log-likelihood gradients within stochastic gradient HMC, but this still results in an inconsistent method.
%
%Stochastic-gradient HMC algorithms~\citep{Welling:2011,chen2014,Ding:2014} performs HMC by computing a stochastic gradient using a subset of the available data with different strategies to maintain a reasonable approximation of the Hamiltonian dynamics.
Hamiltonian ABC~\citep{welling2015} also performs HMC with stochastic gradients but calculates these gradients by using forward simulation. Similarly to stochastic gradient MCMC (and in contrast with \pmhmc), this results in an approximate MCMC which does not preserve the distribution of interest.

Kernel HMC~\citep{Strathmann2015} is another related approach which approximates the gradients by fitting an exponential family model in a reproducing kernel Hilbert space. If the adaptation of the kernel stops after a finite time, or if the adaptation probability decays to zero, then the method can be shown to attain detailed balance, and thus targets the correct distribution of interest. However, the kernel-based approximation gives rise to a bias in the gradients which is difficult to control and there is no guarantee that the trajectories closely follow the ideal HMC. Kernel HMC requires the selection of a kernel and, furthermore, some appropriate approximation thereof, since the computational cost of a full kernel-based approximation grows cubically with the number of MCMC iterations; see \citet{Strathmann2015} for details.

Auxiliary pseudo-marginal methods, also known as \emph{pseudo-marginal slice sampling} (PM-SS;~\citet{MurrayG:2015}) are closely related to \pmhmc in the sense that they target the same extended distribution (given by \eqref{eq:targetpseudomarginal} in the consecutive section) and are therefore also \emph{exact approximations} \citep{andrieu2009pseudo}.
PM-SS, however, relies on alternating the sampling of the parameters given the likelihood estimate and sampling the likelihood estimate given the parameters. In this framework different MCMC methods such as slice sampling~\citep{neal2003} and Metropolis--Hastings can be used to sample from the conditional distributions, typically one always uses \emph{elliptical slice sampling}~\citep{murray2010} for sampling of the likelihood estimates. In comparison our proposed algorithm samples jointly the parameters and the likelihood estimate.

\subsection{Outline of the paper}
\fl{Update this?} \ja{I have updated it.}
%Our paper is organized as follows. In Section~\ref{sec:Hamiltoniandynamics} we present the Hamiltonian dynamics, first for the standard case and after that we introduce the pseudo-marginal Hamiltonian dynamics. In Section~\ref{Section:illustrations} we illustrate the pseudo-marginal Hamiltonian dynamics on the latent variable model. The main contribution of the paper, the pseudo-marginal \hmc algorithm is introduced in Section~\ref{Sec:numericalintegrators}, the theoretical justification of our method is presented in Section~\ref{sec:convergence} with the proofs postponed to the Appendix. Finally we present some numerical results to further prove the usefulness of our algorithm in Section~\ref{sec:applications}.

Our paper is organized as follows. In Section~\ref{sec:Hamiltoniandynamics} we present our algorithm by first introducing in Section~\ref{sec:hmc} the standard Hamiltonian Monte Carlo algorithm. The pseudo-marginal Hamiltonian Monte Carlo algorithm is presented in Section~\ref{sec:pmh} with an illustration on the latent variable model in Section~\ref{Section:illustrations}. The customized numerical integrator used to simulate the Hamiltonian dynamics is presented in Section~\ref{Sec:numericalintegrators} completing the algorithm. The theoretical justification of our algorithm is presented in Section~\ref{sec:convergence} with the proofs postponed to the Appendix. Finally we present some numerical results demonstrating the usefulness of our algorithm in Section~\ref{sec:applications}.

%\section{Hamiltonian dynamics\label{sec:Hamiltoniandynamics}}
\section{Pseudo-marginal Hamiltonian Monte Carlo} \label{sec:Hamiltoniandynamics}

\fl{I have merged the previous sections 2 and 3, since they are not divided by the convergence theory anymore}
\fl{A short "intro" should go here; "First we present some background on Hamiltonian Monte Carlo (HMC). Specifically, we consider the case of targeting the marginal posterior $\pi(\theta)$ using HMC, which results in an ideal but intractable algorithm. We then present pseudo-marginal HMC, an \emph{}}

In this section we present the proposed method.
First we give some background on Hamiltonian Monte Carlo (HMC) and, specifically, we consider the case of targeting the marginal posterior $\pi(\theta)$ using HMC. This results in an ideal but intractable algorithm. We then present pseudo-marginal HMC (\pmhmc), an \emph{exact approximation}, of the marginal HMC.

\subsection{Marginal Hamiltonian Monte Carlo}\label{sec:hmc}

The Hamiltonian formulation of classical mechanics is at the core
of HMC\ methods. Recall that $\theta\in\Theta\subseteq\mathbb{R}^{d}.$
We identify the potential energy as the negative unnormalized log-target
and introduce a momentum variable $\rho\in$ $\mathbb{R}^{d}$ which
defines the kinetic energy $\frac{1}{2}\rho^{T}\rho$ of the system\footnote{For simplicity we assume unit mass. The extension to a general mass matrix is straightforward.}.
The resulting Hamiltonian is given by
\begin{equation}
H_{\mathtt{ex}}(\theta,\rho)=-\log\thinspace p(\theta)-\log\thinspace p(y\mid\theta)+\frac{1}{2}\rho^{T}\rho.\label{eq:Hamiltonianmarginal}
\end{equation}
We associate a probability density on $\mathbb{R}^{d}\times\mathbb{R}^{d}$
to this Hamiltonian through
\begin{align}
\pi(\theta,\rho)  \propto\exp\left(-H_{\mathtt{ex}}(\theta,\rho)\right)
& =\pi\left(\theta\right)\mathcal{N}\left(\rho\mid0_{d},I_{d}\right),
\label{eq:HMCtargetmarginal}
\end{align}
where $\mathcal{N}\left(z\mid\mu,\Sigma\right)$ denotes the normal
density of argument $z$, mean $\mu$ and covariance $\Sigma$.

Assuming that the prior density and likelihood function are continuously
differentiable, the Hamiltonian dynamics corresponds to the equations
of motion
\begin{align}
\frac{\mathrm{d}\theta}{\mathrm{d}t} & =\nabla_{\rho}H_{\mathtt{ex}}=\rho, %\label{eq:HMCdyn1}
&
\frac{\mathrm{d}\rho}{\mathrm{d}t} & =-\nabla_{\theta}H_{\mathtt{ex}}=\nabla_{\theta}\log\thinspace p\left(\theta\right)+\nabla_{\theta}\log\thinspace p(y\mid\theta).\label{eq:HMCdyn}
\end{align}
We will write
\begin{equation}
\varphi\left(\theta,\rho\right)=\left(\begin{array}{c}
\rho\\
\nabla_{\theta}\log\thinspace p\left(\theta\right)+\nabla_{\theta}\log\thinspace p(y\mid\theta)
\end{array}\right).\label{eq:flowhamiltonian}
\end{equation}
A\ key property of this dynamics is that it preserves the Hamiltonian,
i.e. $H_{\mathtt{ex}}(\theta(t),\rho(t))=H_{\mathtt{ex}}(\theta(0),\rho(0))=H_{0}$
for any $t\geq0.$ This enables large moves in the parameter
space to be made by simulating the Hamiltonian dynamics.
However, to sample from the posterior, it is necessary to
explore other level sets of the Hamiltonian; this can be achieved
by periodically updating the momentum $\rho$ according to its marginal
under $\pi$, i.e. $\rho\sim\mathcal{N}\left(0_{d},I_{d}\right)$.

The Hamiltonian dynamics only admits a closed-form solution
in very simple scenarios, e.g., if $\pi\left(\theta\right)$ is normal.
Hence, in practice, one usually needs to resort to a numerical integrator---typically,
the Verlet method, also known as the Leapfrog method, is used due
to its favourable properties in the context of HMC (\citealp[p. 60]{leimkhulermatthews2015};
\citealp[Section 5.2.3.3]{Neal2011}). In particular, this integrator
is symplectic which implies that the Jacobian of the transformation
$\left(\theta(0),\rho(0)\right)\rightarrow\left(\theta(t),\rho(t)\right)$
is unity for any $t>0$. Because of numerical integration errors,
the Hamiltonian is not preserved along the discretized trajectory
but this can be accounted for by an MH\ rejection step. The resulting
HMC\ method is given by the following: at state $\theta:=\thex{0}$,
\emph{(i)} sample the momentum variable $\rhex{0}\sim\mathcal{N}(0_{d},I_{d})$,
\emph{(ii)} simulate approximately the Hamiltonian dynamics over $L$
discrete time steps using a symplectic integrator, yielding $(\thex{L},\rhex{L})$,
and \emph{(iii)} accept $(\thex{L},\rhex{L})$ with probability $1\wedge\pi\left(\thex{L},\rhex{L}\right)/\pi\left(\thex{0},\rhex{0}\right)=1\wedge\exp\left(H_{\mathtt{ex}}(\thex{0},\rhex{0})-H_{\mathtt{ex}}\left(\thex{L},\rhex{L}\right)\right)$.
We refer to \citet{Neal2011} for details and a more comprehensive
introduction.

\subsection{Pseudo-marginal Hamiltonian dynamics}\label{sec:pmh}

When the likelihood is intractable, it is not possible to approximate numerically the Hamiltonian
dynamics (\ref{eq:HMCdyn}), as these integrators
requires evaluating $\nabla_{\theta} \log p(y\mid\theta)$ pointwise.
We will address this difficult by instead considering a Hamiltonian system defined on an extended phase space when the
following assumption holds.
\begin{itemize}[leftmargin=1em]
\item \textbf{Assumption 1}.\label{assumption1:pseudomarginal} There exists
$\left(\theta,\mathbf{u}\right)\mapsto\widehat{p}(y\mid\theta,\mathbf{u})\in\mathbb{R}^{+}$
where $\mathbf{u}\in\mathcal{U}$ and $m\left(\cdot\right)$ a probability
density on $\mathcal{U}$ such that $p(y\mid\theta)=\int\widehat{p}(y\mid\theta,\mathbf{u})m\left(\mathbf{u}\right)\mathrm{d}\mathbf{u}$.
\end{itemize}
Assumption 1 equivalently states that
$\widehat{p}(y\mid\theta,\mathbf{U})$ is a non-negative unbiased
estimate of $p(y\mid\theta)$ when $\mathbf{U}\sim m\left(\cdot\right)$.
This assumption is at the core of pseudo-marginal methods \citep{andrieu2009pseudo,DeligiannidisDP:2015,linliuSloan2000,MurrayG:2015}
which rely on the introduction of an extended target density
\begin{align}
\overline{\pi}\left(\theta,\mathbf{u}\right)  =\pi\left(\theta\right)\frac{\widehat{p}(y\mid\theta,\mathbf{u})}{p(y\mid\theta)}m\left(\mathbf{u}\right)
\propto p\left(\theta\right)\widehat{p}(y\mid\theta,\mathbf{u})m\left(\mathbf{u}\right).
\label{eq:targetpseudomarginal}
\end{align}
This extended target admits $\pi\left(\theta\right)$ as a marginal under Assumption 1. The pseudo-marginal MH\ algorithm
is for example a `standard' MH\ algorithm targeting (\ref{eq:targetpseudomarginal})
using the proposal $q\left(\theta,\theta^{\prime}\right)m\left(\mathbf{u}^{\prime}\right)$
when in state $\left(\theta,\mathbf{u}\right)$, resulting in an acceptance
probability
\begin{equation}
1\wedge\frac{p\left(\theta^{\prime}\right)\widehat{p}(y\mid\theta^{\prime},\mathbf{u}^{\prime})}{p\left(\theta\right)\widehat{p}(y\mid\theta,\mathbf{u})}\frac{q\left(\theta^{\prime},\theta\right)}{q\left(\theta,\theta^{\prime}\right)}.\label{eq:pseudomarginalacceptanceratio}
\end{equation}
Instead of exploring the extended target distribution $\overline{\pi}\left(\theta,\mathbf{u}\right)$
using an MH\ strategy, we will rely here on an HMC\ mechanism. Our
method will use an additional assumption on the distribution of the
auxiliary variables and regularity conditions on the simulated likelihood function $\widehat{p}(y\mid\theta,\mathbf{u})$.
\begin{itemize}[leftmargin=1em]
\item \textbf{Assumption 2}.\label{assumption2:pseudomarginalnormalrv}
$\mathcal{U=}\mathbb{R}^{D}$, $m\left(\mathbf{u}\right)=\mathcal{N}\left(\mathbf{u}\mid0_{D},I_{D}\right)$
and $\left(\theta,\mathbf{u}\right)\longmapsto\widehat{p}(y\mid\theta,\mathbf{u})$
is continuously differentiable and $\left(\theta,\mathbf{u}\right)\longmapsto \log\nabla\widehat{p}(y\mid\theta,\mathbf{u})$
can be evaluated point-wise.
\end{itemize}
This is closely related to the reparametrization trick commonly used in variational inference for unbiased gradient estimation~\citep{KingmaWelling2014}. Our algorithm will leverage the fact that $m\left(\mathbf{u}\right)$ is a normal distribution. Assumptions 1 and 2 will be standing assumptions from now on and
allow us to define the following extended Hamiltonian
\begin{align}
H(\theta,\rho,\mathbf{u},\mathbf{p})
%& = -\log\thinspace p(\theta)-\log\thinspace\widehat{p}(y\mid\theta,\mathbf{u})-\log m\left(\mathbf{u}\right)+\frac{1}{2}\rho^{T}\rho+\frac{1}{2}\mathbf{p}^{T}\mathbf{p}\nonumber \\
=-\log\thinspace p(\theta)-\log\thinspace\widehat{p}(y\mid\theta,\mathbf{u})+\frac{1}{2}\left\{ \rho^{T}\rho+\mathbf{u}^{T}\mathbf{u}+\mathbf{p}^{T}\mathbf{p}\right\} ,\label{eq:extendedHamiltonian}
\end{align}
with a corresponding joint probability density on $\mathbb{R}^{2d+2D}$
\begin{equation}
\overline{\pi}(\theta,\rho,\mathbf{u},\mathbf{p})=\overline{\pi}\left(\theta,\mathbf{u}\right)\mathcal{N}\left(\rho\mid0_{d},I_{d}\right)\mathcal{N}\left(\mathbf{p}\mid0_{D},I_{D}\right),\label{eq:extendedtargetofextendedHamiltonian}
\end{equation}
which also admits $\pi\left(\theta\right)$ as a marginal. Here $\pvar \in \mathbb{R}^{D}$ are momentum variables associated with $\uvar$. The corresponding
equations of motion associated with this extended Hamiltonian are then
given by
\begin{equation}
\frac{d}{dt}\begin{pmatrix}
\theta \\ \rho \\ \mathbf{u} \\ \mathbf{p}
\end{pmatrix} =
\begin{pmatrix}
\rho\\
\nabla_{\theta}\thinspace\log p\left(\theta\right)+\nabla_{\theta}\log\thinspace\widehat{p}(y\mid\theta,\mathbf{u})\\
\mathbf{p}\\
-\mathbf{u}+\nabla_{\mathbf{u}}\log\thinspace\widehat{p}(y\mid\theta,\mathbf{u})
\end{pmatrix}
=:
\widehat{\Psi}\left(\theta,\rho,\mathbf{u},\mathbf{p}\right).
\label{eq:pseudoHMCflow}
\end{equation}

Compared to (\ref{eq:HMCdyn}), the intractable log-likelihood gradient
$\nabla_{\theta}\thinspace\log p(y\mid\theta)$ appearing in (\ref{eq:HMCdyn})
has now been replaced by the gradient $\nabla_{\theta}\thinspace\log\widehat{p}(y\mid\theta,\mathbf{u})$
of the log-simulated likelihood $\widehat{p}(y\mid\theta,\mathbf{u})$
where $\mathbf{u}$ evolves according to the third and fourth rows of (\ref{eq:pseudoHMCflow}). %(\ref{eq:HMCextendeddyn3})-(\ref{eq:HMCextendeddyn4}).

\begin{remark}
The normality assumption is not restrictive as we
can always generate a uniform random variate from a normal one using
the cumulative distribution function of a normal. This assumption has
also been used by \citet{DeligiannidisDP:2015} for the correlated pseudo-marginal MH\ method and \citet{MurrayG:2015} for pseudo-marginal slice
sampling. The assumed regularity of the simulated likelihood function $\widehat{p}(y\mid\theta,\mathbf{u})$
is necessary to implement the pseudo-marginal HMC but, unfortunately, limits its range
of applications. For example, in a state-space
model context, the likelihood is usually estimated using a particle
filter as in \citet{andrieudoucetholenstein2010} but this results in a\ discontinuous function $\left(\theta,\mathbf{u}\right)\longmapsto\widehat{p}(y\mid\theta,\mathbf{u})$. 	
 \end{remark}

%\subsection{Illustration on latent variable models\label{Section:illustrations}}

%We revisit here the examples described in the introduction.

%\noindent
%\vspace{1ex}%
%\textbf{Example: Illustration on latent variable models.}
\subsection{Illustration on latent variable models}\label{Section:illustrations}%
Consider the latent variable model described by (\ref{eq:independentlatentvariablemodels}) and (\ref{eq:integrallikelihoodpaneldata}).
In this scenario, the intractable likelihood can be unbiasedly estimated
using importance sampling. We introduce an importance density $q_{\theta}(x_{k}\mid y_{k})$
for the latent variable $X_{k}$ which we assume can be simulated
using $X_{k}=\gamma_{k}(\theta,V)$ where $\gamma_{k}:\Theta\times\mathbb{R}^{p}\rightarrow\mathbb{R}^{n}\ $is
a deterministic map and $V\sim\mathcal{N}\left(0_{p},I_{p}\right)$.\ We
can then approximate the likelihood, using $N$ samples for each $k$, through
\begin{align}
\widehat{p}(y_{1:T}\mid\theta,\mathbf{U}) &= { \prod _{k=1}^{T}}\,\widehat{p}(y_{k}\mid\theta,\mathbf{U}_{k}),
&
&\text{where}
&
\widehat{p}(y_{k}\mid\theta,\mathbf{U}_{k})&=\frac{1}{N}\sum_{i=1}^{N}\omega_{\theta}\left(y_k, \mathbf{U}_{k,i}\right),
\label{eq:likelihoodestimator}
%\label{eq:ISestimatesindividuallikelihoodterms}
\end{align}
with $\mathbf{U}_{k}:=\left(\mathbf{U}_{k,1},\ldots,\mathbf{U}_{k,N}\right)$
and
%\begin{equation}
\(
\omega_{\theta}\left(y_{k},\mathbf{U}_{k,i}\right)=\frac{g_{\theta}(y_{k}\mid X_{k,i})f_{\theta}(X_{k,i})}{q_{\theta}(X_{k,i}\mid y_{k})}, %\label{eq:ISweights}
\)
%\end{equation}
where $X_{k,i}=\gamma_{k}(\theta,\mathbf{U}_{k,i})$ and $\mathbf{U}_{k,i}\overset{\text{i.i.d.}}{\sim}\mathcal{N}\left(0_{p},I_{p}\right)$.
We thus have $D=TNp$ in this scenario and
\begin{align}
\nabla_{\theta}\log\thinspace\widehat{p}(y_{1:T}\mid\theta,\mathbf{U})&=\sum_{k=1}^{T}\sum_{i=1}^{N}\frac{\omega_{\theta}\left(y_{k},\mathbf{U}_{k,i}\right)}{\sum_{j=1}^{N}\omega_{\theta}\left(y_{k},\mathbf{U}_{k,j}\right)}\nabla_{\theta}\log\omega_{\theta}\left(y_{k},\mathbf{U}_{k,i}\right),\label{eq:simulatedgradienttheta}
\\
\nabla_{\mathbf{u}_{k,i}}\log\thinspace\widehat{p}(y_{1:T}\mid\theta,\mathbf{U})&=\frac{\omega_{\theta}\left(y_{k},\mathbf{U}_{k,i}\right)}{\sum_{j=1}^{N}\omega_{\theta}\left(y_{k},\mathbf{U}_{k,j}\right)}\nabla_{\mathbf{u}_{k,i}}\log\omega_{\theta}\left(y_{k},\mathbf{U}_{k,i}\right).\label{eq:simulatedgradientu}
\end{align}
The pseudo-marginal MH\ algorithm can mix very poorly if the relative variance of the likelihood estimator is large; e.g. if $N=1$ in  (\ref{eq:likelihoodestimator}). In the pseudo-marginal Hamiltonian dynamics context, the case $N=1$ corresponds to an Hamiltonian dynamics on a re-parameterization of the original joint model and can work well
%in the absence of multimodality;
for simple targets;
see e.g. \citep{betancourt2015}.
Thus, we expect pseudo-marginal HMC to be much less sensitive to the choice of $N$ than pseudo-marginal MH; see Section~\ref{sec:discussion} for empirical results.

The construction above is based on a standard importance sampling estimator of the likelihood, but most sophisticated estimators could be used. For instance,~\citet{kloster2018} have recently investigated the use of \emph{efficient importance sampling} \citep{richard:2007} to approximate the likelihood within a \pmhmc algorithm.

%\section{Pseudo-marginal Hamiltonian Monte Carlo\label{Sec:numericalintegrators}}
\subsection{Numerical integration via operator splitting\label{Sec:numericalintegrators}}
The pseudo-marginal Hamiltonian dynamics (\ref{eq:pseudoHMCflow})
can not in general be solved analytically and, as usual, we therefore
need to make use of a numerical integrator. The standard choice in
HMC is to use the Verlet scheme \citep[Section 2.2]{leimkhulermatthews2015}
which is a symplectic integrator of order $O(h^{2})$, where $h$
is the integration step-size. However, the error of the Verlet integrator
will also depend on the dimension of the system. For the pseudo-marginal
target density (\ref{eq:extendedtargetofextendedHamiltonian}), we
therefore need to take the effect of the $D$-dimensional auxiliary
variable $\mathbf{u}$ into account. For instance, in the context
of importance-sampling-based pseudo-marginal HMC\ for latent variable
models
discussed above
%of Section \ref{Section:illustrations},
 we have $D=TNp$,
i.e., the dimension of the extended target increases linearly with
the number of importance samples $N$. This is an apparent problem---by increasing
$N$ we expect to obtain solution trajectories closer to those of
the true marginal Hamiltonian system. However we also need to integrate numerically an ordinary different equation of dimension increasing with $N$ so one might fear that the overall numerical integration error increases.
%Nevertheless we manage in the next section prove both that the $(\theta,\rho)$ marginal of the trajectory in the \pmhmc algorithm will converge to the ideal \hmc trajectory. Also we provide results of the acceptance probability and show that it will converge to the acceptance probability corresponding to the ideal \hmc algorithm. That is there is no problem with the increasing dimension that is connected to increasing $N$.

However, it is possible to circumvent this problem by making
use of a splitting technique which exploits the stucture of the extended target, see~(\citealp{Beskos:2011}; \citealp[Section 2.4.1]{leimkhulermatthews2015};
\citealp[Section 5.5.1]{Neal2011}). The idea is to split the Hamiltonian $H$ defined
in (\ref{eq:extendedtargetofextendedHamiltonian}) into two components
$H=A+B$, where
\begin{align}
A(\rho,\mathbf{u},\mathbf{p}) & :=\frac{1}{2}\left\{ \rho^{T}\rho+\mathbf{u}^{T}\mathbf{u}+\mathbf{p}^{T}\mathbf{p}\right\}, &
B(\theta,\mathbf{u}) & :=-\log\thinspace p(\theta)-\log\thinspace\widehat{p}(y_{1:T}\mid\theta,\mathbf{u}).
%\label{eq:splitting:split-H1-def}
%\label{eq:splitting:split-H2-def}
\end{align}
The Hamiltonian systems for $A$ and $B$ can both be integrated analytically.
Indeed, if we define the mapping $\Phi_{h}^{A}:\mathbb{R}^{2d+2D}\mapsto\mathbb{R}^{2d+2D}$
as the solution to the dynamical system with Hamiltonian $A$ simulated for $h$ units of time from a given initial condition, we have the explicit solution
\fl{Would it make sense to explicitly write out what the Hamiltonian dynamics for systems A and B are? One reviewer for the NIPS submission was confused about our claim that we can integrate A and B exactly.} \ja{I think that would mostly confuse readers.}

\fl{We shouldn't have brackets here; $h$ is "real" time not discrete time steps} \ja{fixed}
\begin{equation}
\Phi_{h}^{A}:\begin{cases}
\thexp{h} =\thexp{0} + h \rhexp{0}, \\
\rhexp{h} = \rhexp{0}, \\
\uexp{h} = \pexp{0} \sin(h) + \uexp{0} \cos(h), \\
\pexp{h} = \pexp{0} \cos(h) - \uexp{0} \sin(h).
\end{cases} \label{eq:PhiA}
\end{equation}

Similarly for system $B$ we define the mapping $\Phi^{B}_h : \rset^{2d + 2D} \mapsto \rset^{2d + 2D}$ and get the solution,

\begin{equation}
\Phi_{h}^{B}:\begin{cases}
\thexp{h} = \thexp{0},\\
\rhexp{h} = \rhexp{0} + h \nabla_{\theta}\{ \log p(\theta) + \log \widehat{p}(y_{1:T} \mid \theta, \uexp{0} ) \}_{|\theta = \thexp{0}}, \\
\uexp{h} = \uexp{0}, \\
\pexp{h} = \pexp{0} + h \nabla_{\uvar} \log \widehat{p}(y_{1:T} \mid \thexp{0}, \uvar)_{| \uvar = \uexp{0}}
\end{cases}\label{eq:PhiB}
\end{equation}

Let the integration
time be given as $hL$, where $h$ is the step-size and $L$ the number of integration steps.
To approximate the solution %$\widehat{\Phi}_{hL}$
to the original
system associated to the vector field $\widehat{\Psi}$ in \eqref{eq:pseudoHMCflow},
we then use a
%numerical integrator defined as follows (known as
symmetric Strang splitting (see, e.g., \citet[p. 108]{leimkhulermatthews2015})
defined as
$\widehat{\Phi}_{hL}=\{\Phi_{h/2}^A\circ\Phi_{h}^B\circ\Phi_{h/2}^A\}^{\circ L}$.
In practice we combine consecutive half-steps of the integration of system $A$ for numerical efficiency, similarly to what is often done for the standard Verlet integrator, \ie, we use
\begin{equation}
	\widehat{\Phi}_{hL} = \Phi_{h/2}^A\circ \{ \Phi_{h}^B\circ\Phi_{h}^A\}^{\circ L-1} \circ\Phi_{h}^B\circ\Phi_{h/2}^A. \label{eq:integrator:in:practice}
\end{equation}
In Appendix~\ref{app:int} the explicit update equations corresponding to a full step of $\widehat{\Phi}_h$ are given.

Using this integration method we can prove (see the next section) both that the $(\theta, \rho)$-trajectory of the \pmhmc algorithm converges to the ideal \hmc trajectory, and also that the acceptance probability of the \pmhmc algorithm converges to that of the ideal \hmc algorithm. This shows that, when using the aforementioned integration technique, the increase in dimension which happens with increased $N$ does not pose any problem with respect to the convergence of the algorithm to its idealized counterpart.

An alternative method which also exploits the structure of the extended Hamiltonian and that could be used in our setting is the exponential integration technique of \cite{chao2015exponential}. In simulations we found the two integrators to perform similarly and we focus on the splitting technique for simplicity.

The proposed pseudo-marginal HMC algorithm is summarized in Algorithm~\ref{alg:pmhmc}. As this algorithm simulates by design a Markov chain of invariant distribution $\overline{\pi}(\theta,\rho,\mathbf{u},\mathbf{p})$ defined in \eqref{eq:extendedtargetofextendedHamiltonian}, it samples asymptotically (in the number of iterations) from its marginal $\pi\left(\theta\right)$ under ergodicity conditions.

\begin{algorithm}
\protect\caption{Pseudo-marginal HMC (one iteration)}

\fl{This is not a very detailed algorithm. Would it be better/more helpful to write it out more explicitly?} \ja{Not sure how that would be done in a good way, added the numerical integrator as an equation and referenced that. }
\label{alg:pmhmc} Let $(\theta,\mathbf{u})$ be the current state
of the Markov chain. Do:
\begin{enumerate}
	\item Sample auxiliary variables $\rho\sim\mathcal{N}\left(0_{d},I_{d}\right)$ and $\mathbf{p}\sim\mathcal{N}(0_D,I_{D})$.
	\item Compute $(\theta',\rho',\mathbf{u}',\mathbf{p}')=\widehat{\Phi}_{hL}(\theta,\rho,\mathbf{u},\mathbf{p})$ using the numerical integrator~\eqref{eq:integrator:in:practice}.
	\item Accept $(\theta',\mathbf{u}')$ with probability
	\(
	1 \wedge \exp\left(H(\theta,\rho,\mathbf{u},\mathbf{p})-H(\theta',\rho',\mathbf{u}',\mathbf{p}')\right),
	\)
\end{enumerate}
\end{algorithm}

\section{Convergence results}\label{sec:convergence}

In this section we establish the convergence (under suitable assumptions) of the \pmhmc in the sense that it will converge as $N$ increases towards the ideal marginal \hmc algorithm. We begin by studying the numerical integrator $\widehat{\Phi}_{hL}$ and wish to limit the error in the first two components when comparing with the results from the ideal \hmc algorithm.

%We start by only focusing on the numerical integrator, showing that as $N$ increase the marginal of interest will converge to the values given by the \hmc method using the same numerical integrator.

For the rest of this section we will adopt the following notation: we use $\parvecest{\ell} = (\thest{\ell}, \rhest{\ell}, \uest{\ell}, \pest{\ell})^\top$ to denote the results of running the \pmhmc, and $\parvec{\ell}$ the results of running the ideal \hmc algorithm (which is intractable), for $\ell$ iterations of the corresponding numerical integrator. For the ${\bf{u}}$ and ${\bf{p}}$ variables we use the notation $\uest{\ell}_{k,i}$ to denote position $i$ in the vector associated with the observation $y_k$, this is consistent with the notation used in Section~\ref{Section:illustrations}. Further we use the notation $\| \cdot \|$ for the Euclidean norm, $\convd$ for convergence in distribution, and $\convp$ for convergence in probability. We focus here on the setting where we use the latent variable model~\eqref{eq:independentlatentvariablemodels}--\eqref{eq:integrallikelihoodpaneldata} and the importance sampling estimator~\eqref{eq:likelihoodestimator}--\eqref{eq:simulatedgradientu}. The proofs of these results are postponed to the Appendix.

We also present the following assumption on the weight function that will be needed for the proofs.
\begin{itemize}[leftmargin=1em] %[noitemsep, topsep=0pt]
	\item \textbf{Assumption 3}. \label{assumption:cont}
	The importance weight $\omega_{\theta}(y, \uvar)$ defined in Section~\ref{Section:illustrations} satisfies:
	\begin{itemize}[topsep=0pt]
		\item $(\theta, \uvar) \to \nabla_{\uvar} \log \omega_{\theta}(y,\uvar)$ is Lipschitz with constant $M$ uniformly in $y$,
		\item $(\theta, \uvar) \to \omega_{\theta}(y, \uvar)$ is Lipschitz with constant $D$ uniformly in $y$,
		\item there exists constants $0 < \wgtlow < \wgtup < \infty$ such that $\wgtlow < \omega_{\theta}(y,\uvar) < \wgtup$,
		\item $\norm{\nabla_{\uvar} \log \omega_{\theta}(y, \uvar)}$ is bounded from above by $C < \infty$.
	\end{itemize}	
\end{itemize}

This assumption is quite restrictive and does not hold for most practical problems. Nonetheless we have chosen to use it to keep the following theoretical analysis simple and to the point. In the simulations below we look at models that violate this assumption and show that the algorithm still performs as expected. We believe that it is possible to relax these conditions but that is beyond the scope of this paper.

\begin{proposition}\label{prop:difference}
Let $(\thex{L}, \rhex{L})$ be the value associated with the ideal \hmc dynamics and $(\thest{L}, \rhest{L})$ be the values associated to the $(\theta,\rho)$-marginal of the \pmhmc dynamics after $L$ steps of the numerical integrator using the Strang splitting with step-size $h$. Furthermore, let both of the processes start in the same point, i.e.~$(\thex{0},\rhex{0}) = (\thest{0},\rhest{0})$.

	Assume that Assumption 3 holds and that $\nabla_{\theta} \log p(\theta \mid y)$ is Lipschitz with constant $L_0 < \infty$. Then there exists a constant $\lip < \infty$, which does not depend on $N$ and $L$, such that for any $L \geq 1$
	and any choice of initial values $(\thex{0},\rhex{0}) = (\thest{0},\rhest{0})$, and $(\uest{0}, \pest{0})$,
	\begin{align*}
	& \left\|\parmatsmallest{L} - \parmatsmall{L} \right\|  \leq h^2 \sqrt{\frac{h^2}{4} + 1} \\
	& \hspace{.5 cm} \times \sum_{\ell = 0}^{L - 1} (1 + h \lip)^{L - (\ell + 1)}  \left\| \nabla_{\theta} \log \left( \frac{\widehat{p}(y \mid \theta, \pest{\ell} \sin(\frac{h}{2}) + \uest{\ell}\cos(\frac{h}{2}))}{p(y \mid \theta)} \right)_{\mid \theta = \thest{\ell} + \frac{h}{2}\rhest{\ell}} \right\|.
	\end{align*}
	\end{proposition}

	\fl{I moved the second part of the proposition out}

	By taking the expected value of both sides conditioned on the initial values of $(\thex{0}, \rhex{0}) = (\thest{0},\rhest{0})$ and using Jensen's inequality we get as an immediate corollary that
	\begin{align*}
	& \E\left[ \left\|\parmatsmallest{L} - \parmatsmall{L} \right\|\right] \leq h^2 \sqrt{\frac{h^2}{4} + 1} \\
	& \hspace{.5 cm}\times \sum_{\ell = 0}^{L - 1} (1 + h \lip)^{L - (\ell + 1)} \E\left[ \left\| \nabla_{\theta} \log \left( \frac{\widehat{p}(y \mid \theta, \pest{\ell} \sin(\frac{h}{2}) + \uest{\ell}\cos(\frac{h}{2}))}{p(y \mid \theta)} \right)_{\mid \theta = \thest{\ell} + \frac{h}{2}\rhest{\ell}} \right\|^2  \right]^{1/2}.
	\end{align*}
	%By using Jensen's inequality  we observe that this
	That is, the upper bound is directly related to the second moment of the error in the log-likelihood gradient $\nabla_{\theta}\log(\widehat{p}(y \mid \theta, \uvar) / p(y \mid \theta))$. The next result establishes a CLT for this error as $N$ grows in two scenarios. First we study the behavior at stationarity, that is when $\uest{0} \sim \bar{\pi}(\cdot \mid \theta)$. Second we show that a similar CLT holds when $\uest{0} \sim \N(0_D,I_D)$, that is at initialization of the algorithm.

	In the following we will assume that $\theta$ is scalar for notational simplicity. In the multivariate case the results should be interpreted to hold component-wise. We will by $\E_{\N}$ denote the expected value under the standard normal distribution of appropriate dimension.

	\begin{proposition}\label{prop:clt}
	Suppose that Assumption 3 holds. Let $\varpi_{\theta}(y,\uvar) \eqdef \omega_{\theta}(y, \uvar)/p(y \mid \theta)$ and assume that
	\fl{Why is there a subscript $i$ on $u$ in the next expression?} \ja{error, fixed.}
	 $\uvar \to \nabla_{\theta} \varpi_{\theta}(y, \uvar)$ is continuous and that $\norm{ \nabla_\theta \varpi_{\theta}(y, \uvar)}$ is bounded from above by a constant $E < \infty$. Further assume that $\E_{\N}[\varpi^2_{\theta}(y_k, \uvar)] < \infty$ and $\E_{\N}[\{\nabla_{\theta}\varpi(y_k, \uvar)\}^{2}] < \infty$, there exists a function $g_k(\uvar)$ which may depend on $\theta$ and $y_k$ such that $|\nabla_{\theta}\varpi(y_k, \uvar)| < g_k(\uvar)$ and $\E_{\N}[g_k(\uvar)] < \infty$ and $\E_{\N}[\varpi_{\theta}(y_k, \uvar) |\nabla_{\theta}\log\varpi_{\theta}(y_k, \uvar)|] < \infty$ for all $k = 1, \ldots, T$.

	Then, for any $\ell\geq 1$, the following CLT holds when $\uest{0} \sim \bar{\pi}( \cdot \mid \theta)$ and $\pest{0} \sim \N(0_D, I_D)$:
	\begin{align*}
	\sqrt{N} \nabla_{\theta} \log \frac{\widehat{p}(y_{1:T} \mid \theta, \pest{\ell}\sin(\tfrac{h}{2}) + \uest{\ell}\cos(\tfrac{h}{2}))}{p(y \mid \theta)} \convd \N(0, \sigma^2(y_{1:T}, \theta)), \quad \text{as } N \to \infty,
	\end{align*} \fl{Change from $i$ to $\ell$?} \ja{fixed}
	where the variance is given by
	\begin{equation*}
	\sigma^2(y_{1:T}, \theta) = \sum_{k=1}^{\T} \E_{\N}[\{\nabla_{\theta} \varpi_{\theta}(y_{k}, \vvar_{k,1})\}^2].% \quad \vvar_{k,1} \sim \N(0_p, I_p).
	\end{equation*}
	The same CLT also holds when $\uest{0} \sim \N(0_D,I_D)$.
	\end{proposition}

	We now look at the acceptance probability of \pmhmc. As we have shown above, the trajectory of the $(\theta, \rho)$-marginal of the extended space used in \pmhmc converges towards the trajectory of the ideal \hmc algorithm as $N$ increases. Thus we also expect that the acceptance probability of the \pmhmc algorithm converges towards the acceptance probability of the \hmc algorithm at equilibrium. This is established in the following proposition.

	\begin{proposition} \label{prop:acc}
	\fl{The first sentence here is not very informative, since this is the same for all results presented above as well (so it's a bit strange to point it out here). However, it would be useful to say what is assumed about the initial condition for this to hold. Arbitrary? Stationarity? What is the distribution of the auxiliary variables?} \ja{added the information}
	%Let $(\thest{\ell}, \rhest{\ell}, \uest{\ell}, \pest{\ell})$ for $\ell = 0, \ldots, L$ be the values associated with running the \pmhmc algorithm using the numeric integrator $\widehat{\Phi}_{h\ell}$.

	Let Assumption 3 hold, for any $\thest{0}$ assume that $\uest{0} \sim \bar{\pi}(\, \cdot \mid \thest{0})$, $\rhest{0} \sim \N(0_d, I_d)$, and $\pest{0} \sim \N(0_D, I_D)$, we then have that
	\begin{align*}
	&H(\thest{0},\rhest{0},\uest{0},\pest{0}) - H(\thest{L}, \rhest{L}, \uest{L}, \pest{L}) \\
	&\hspace{4 cm}\convp H_{\mathtt{ex}}(\thest{0}, \rhest{0}) - H_{\mathtt{ex}}(\thest{L},\rhest{L}), \quad \text{as } N \to \infty.
	\end{align*}
	Here $H_{\mathtt{ex}}$ is the Hamiltonian associated with the ideal algorithm given in~\eqref{eq:Hamiltonianmarginal}.
	\end{proposition}

As mentioned in Section~\ref{Sec:numericalintegrators}, we again note that as $N$ increases the dimension of the ordinary differential equations one needs to approximate numerically. At a glance this might seem problematic, since the integration error typically increases with the dimension, but in this section we have established that the increase of dimension with $N$ does not pose a problem to our algorithm. In fact, for any step-size $h$ and number of integration steps $L$ the $(\theta,\rho)$-marginal of the \pmhmc algorithm will converge towards the values of the ideal \hmc algorithm and the acceptance probability (directly related to the difference in the Hamiltonian values) converges to the acceptance probability of the ideal \hmc algorithm.

\section{Numerical Illustrations}\label{sec:applications}%
We illustrate the proposed \pmhmc method on three
synthetic examples. Additional details and results are given in the \suppmat.

For simplicity, we focus on the case when standard importance-sampling-based estimators of the likelihood are used to construct the extended Hamiltonian system. However, more efficient estimators can be used and they will intuitively improve the performance of \pmhmc. One illustration of this is given by \citet{kloster2018}, who consider the use of \emph{efficient importance sampling} in the context of \pmhmc for inference in dynamical systems. We refer to this article for additional numerical illustrations of \pmhmc.

%For more examples we refer the reader to the paper by~\citet{kloster2018}.

\subsection{Gaussian model}\label{sec:gauss}
\begin{figure}[ptb]
\centering
\includegraphics[width=0.85\linewidth]{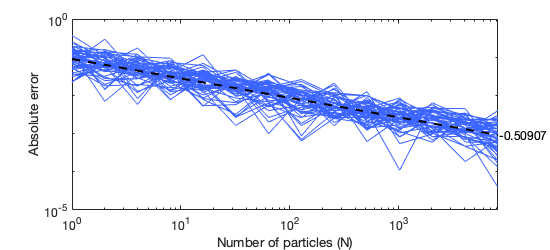}
\caption{Aboslute error $\max_{\ell \in [0,10]}|\thex{\ell} - \thest{\ell}|$ for different initial conditions. The slope of the linear fit, shown as a dashed line, is $-0.509$. \fl{Slope does not correspond to value in figure. Which is it?} \ja{Plot is correct, fixed.}}
\label{fig:conv:traj}
\end{figure}

We consider first the following Gaussian model where $X_k \mid \theta \overset{\text{i.i.d.}}{\sim} \N(\theta, \sigma_X^2)$ and $Y_k \mid (X_k = x_k), \theta \sim \N(x_k, \sigma_Y^2)$ and assign a Gaussian prior on the parameter, $\theta \sim \N(\mu_{\theta}, \sigma_{\theta}^2)$. We choose this model as a first illustration since the true posterior will be a Gaussian. This allows us to run the ideal marginal \hmc algorithm targeting $p(\theta\mid y_{1:T})$ and we can then compare our \pmhmc algorithm with the ideal \hmc algorithm. We simulate $T=30$ i.i.d.~observations using $\mu_{\theta} = 0, \sigma^2_{\theta} = 10, \sigma_x^2 = 0.1,$ and $\sigma_y^2 = 1$. We apply the \pmhmc algorithm with $N = 2^i$ for $i$ from $0$ to $13$. First we check the convergence of the trajectories from the numerical integrator $\hat{\Phi}_{hL}$ towards the trajectories for the ideal \hmc algorithm. We do this by using $h = 0.1$ and $L = 10$ and look at the maximal position error over the integration period. We run the algorithm $50$ times using different starting values and look at how the maximal error depends on the choice of $N$. The results, which be seen in Figure~\ref{fig:conv:traj}, imply a $\sqrt{N}$ convergence rate of the maximal error.

\begin{figure}[ptb]
\centering
	\includegraphics[width=0.85\linewidth]{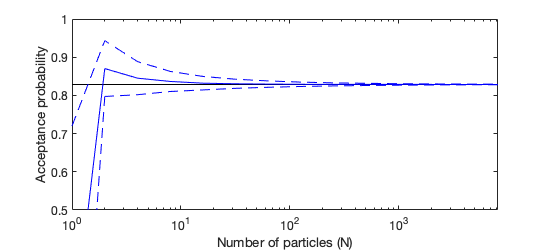}%
	\caption{Acceptance probability of the \pmhmc algorithm as a function of $N$ in a Gaussian example. The dashed lines correspond to a 95\% confidence interval on the acceptance probability. The horizontal line indicates the acceptance probability of the ideal HMC algorithm. {The axes are cropped for clarity and only the values for $N \geq 2$ are shown.} For $N = 1$ the mean is $0.16$.}
	\label{fig:acc:prob}
	\end{figure}

Next we look at the convergence of the acceptance probability as a function of $N$. The results are based on \thsnd{2} independent runs of the algorithm using $h=0.2$ and $L=10$. All of the runs are initialized using the same values of $\thest{0}$ and $\rhest{0}$ while $\uest{0}$ and $\pest{0}$ are randomized for each run. This is done so that we can compare with the acceptance probability given by the ideal \hmc algorithm for the same initial values. The results can be seen in Figure~\ref{fig:acc:prob} where we present the mean of our runs together with 95\% confidence interval. We can clearly see that the acceptance probability converges towards the true value as $N$ increases.

\fl{Can we get the figures as pdf instead? :) Also, I'd add additional tick labels on the y-axis in Fig 1} \ja{you can get eps figures (matlab to pdf is not really working for me). Added more ticks}

	\subsection{Diffraction model}\label{sec:sinc2}
	Consider a hierarchical model of the form
%\begin{align*}
\(X_k \mid \theta \overset{\text{i.i.d.}}{\sim} \N(\mu, \sigma^2)\) and
\(
Y_k \mid (X_k=x_k), \theta \sim g(\cdot \mid x_k, \lambda),
\)
%\end{align*}
with $\theta = (\mu, \log(\sigma), \log(\lambda))$, where the observation density is modelled as a diffraction intensity:
%\begin{align*}
\(
g(y \mid x, \lambda) = (\lambda\pi)^{-1} \sinc^2( \lambda^{-1} (y-x)) .
\)
%\end{align*}
We simulate $\T=100$ \iid observations using $\mu=1$, $\sigma=1$, and $\lambda=0.1$.
%A normal $\N(0,10^2)$ prior is used for each component of $\theta$.
We apply the
%pseudo-marginal HMC (henceforth, PM-HMC)
PM-HMC
with $N$ ranging from 16 to 256, using a non-centered parameterization and the prior as importance density,
as well as a standard HMC working on the joint space using the same non-centered parameterization\footnote{The standard HMC, here, is equivalent to PM-HMC with $N=1$. Specifically, it operates on the joint, non-centered $(\theta,\mathbf{u})$ space, and it uses the splitting integrator described in Section~\ref{Sec:numericalintegrators}.}.
For further comparison we also ran, \emph{(i)} a standard pseudo-marginal MH algorithm using $N=512$ and $N=1024$ (smaller $N$ resulted in very sticky behavior), \emph{(ii)} the pseudo-marginal slice sampler by \cite{MurrayG:2015} with $N$ ranging from 16 to 256, and \emph{(iii)} a Gibbs sampler akin to the Particle Gibbs algorithm by \cite{andrieudoucetholenstein2010} (using independent conditional importance sampling kernels to update the latent variables).

%, discarding the first $\thsnd{10}$ as burn-in.

\begin{figure}[ptb]
\centering
	\includegraphics[width=0.33\linewidth]{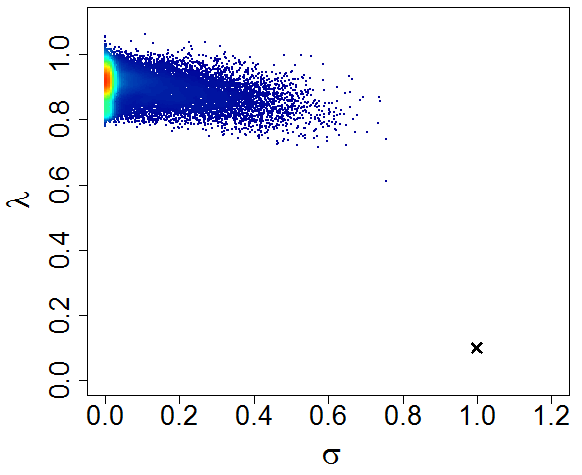}%
	\includegraphics[width=0.33\linewidth]{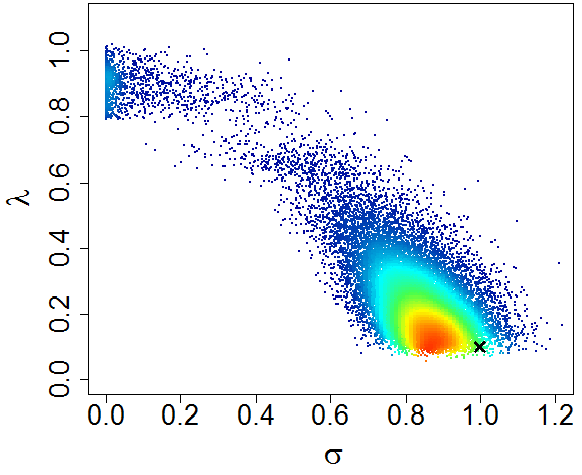}%
	\includegraphics[width=0.33\linewidth]{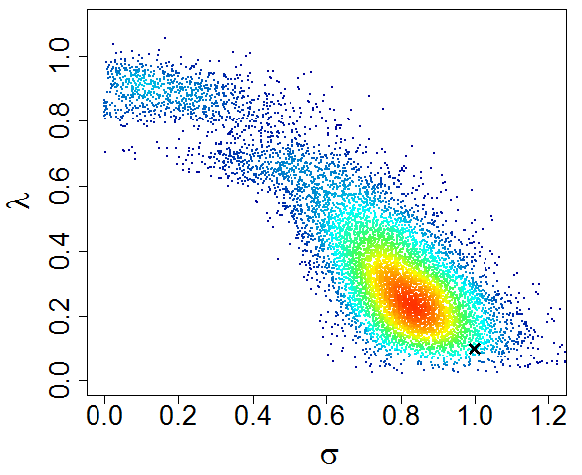}%
	\caption{Scatter plots for $(\sigma, \lambda)$ for
	standard HMC using a non-centered parameterisation (left),
	PM-HMC with $N=16$ (mid), and
	pseudo-marginal slice sampling with $N=128$ (right). The black {\bf x} corresponds to the variables used to generate the data.}
	\label{fig:res:sinc2:scatter}
	\end{figure}

Figure~\ref{fig:res:sinc2:scatter} shows scatter plots
from $\thsnd{40}$ iterations (after a burn-in of $\thsnd{10}$)
for the parameters $(\sigma, \lambda)$ for standard \hmc, \pmhmc ($N=16$),
and pseudo-marginal slice sampling ($N=128$, smaller values gave very poor mixing). Results for the remaining methods/settings,
including the pseudo-marginal MH method and the Particle Gibbs sampler which both performed very poorly, are given in the \suppmat.
For the pseudo-marginal slice sampling we ran a SS+MH algorithm, where we use a random walk for the $\theta$ variables and elliptical slice sampling for the $\uvar$ variables.
\fl{Add details on how PMSS is implemented} \ja{added}

It is clear that \hmc fails to explore the distribution due to the raggedness of the joint posterior. \pmhmc effectively marginalizes out the multimodal latent variables and effectively samples the marginal posterior which, albeit still bimodal, is much smoother than the joint posterior. Pseudo-marginal slice sampling also does well in this example and we found the two methods to be comparable in performance when normalized by computational cost.\footnote{We do not report effective sample size estimates, as we found these to be very noisy and misleading for this multimodal distribution. Indeed, the ``best'' effective sample size when normalized by computational cost was obtained for the Particle Gibbs sampler which, by visual inspection, completely failed to converge to the correct posterior.}
%The posterior is clearly multimodal, as identified by \pmhmc, whereas standard \hmc gets stuck in a mode far from the data generating parameter values.

	\subsection{Generalized linear mixed model}\label{sec:glmm}
	Next, we consider inference in a generalized linear mixed model (GLMM), see \eg \cite{zhao2006}, with a logistic link function:
%\begin{align*}
%&Y_{ij} \sim \text{Bernoulli}(p_{ij}), \\%& &\text{where} &
%&\text{logit}(p_{ij}) = B_i + X_{ij}^\+ \beta,
%\end{align*}
$Y_{ij} \sim \text{Bernoulli}(p_{ij})$, where $\text{logit}(p_{ij}) = X_i + Z_{ij}^\+ \beta$ for $i=\range{1}{\T}$, $j=\range{1}{n_i}$.
Here, $Y_{ij}$ represent the $j$th observation for the $i$th ``subject'',
$Z_{ij}$ is a covariate of dimension $p$, $\beta$ is a vector of fixed effects
and $X_i$ is a random effect for subject~$i$.
It has been recognized \citep{burda2008,komarek2008} that it is often beneficial to allow for non-Gaussianity in the random effects. For instance, multi-modality can arise as an effect of under-modelling, when effects not accounted for by the covariates result in a clustering of the subjects. To accommodate this we assume $X_i$ to be distributed according to a Gaussian mixture:
%\begin{align*}
\(
X_i \overset{\text{i.i.d.}}{\sim} \sum_{j=1}^K w_j \N(\mu_j, \lambda_j^{-1}).
\)
%\end{align*}
For simplicity we fix $K=2$ for this illustration. The parameters of the model are thus $\beta$, $\{\mu_j, \lambda_j\}_{j=1}^2$, and $w_1$ (as $w_2 = 1-w_1$), with $X_{1:\T}$ being latent variables. We use the parameterisation
\begin{align*}
\theta = (\beta^\+, \mu_1, \mu_2, \log(\lambda_1), \log(\lambda_2), \text{logit}(w_1) )^\+ \in \reals^{p+5}.
\end{align*}
%
%and assign a $\N(0, 100)$ prior to each component of $\theta$.
%\note{beta has a different prior?}
%
We used a simulated data set with $\T=500$ and $n_i\equiv 6$, thus a total of 3\,000 data points, with $\mu_1=0$, $\mu_2=3$, $\lambda_1=10$, $\lambda_2=3$.
We set $p=8$ and generate $\beta$ as well the covariates $Z_{ij}$ from standard normal distributions (see the \suppmat for further details on the simulation setup).

\fl{Add details on how PMSS is implemented} \ja{added}

\begin{figure}[ptb]
\centering
	\includegraphics[width=0.33\linewidth]{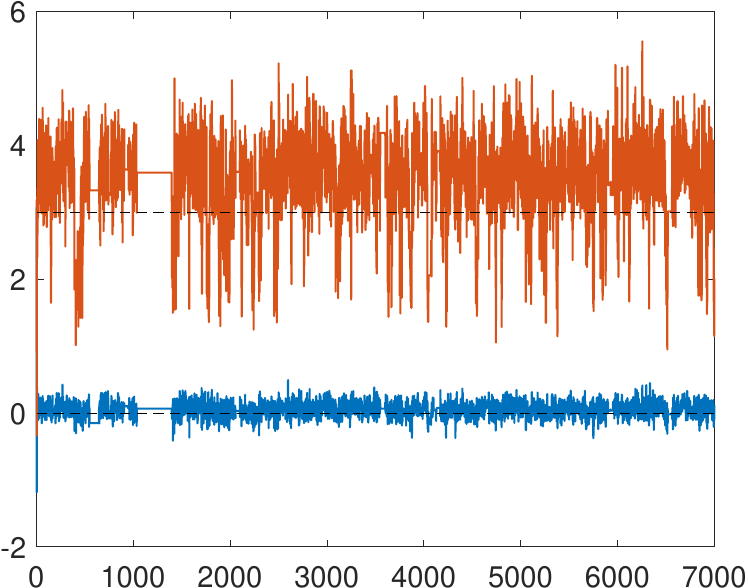}%
	\includegraphics[width=0.33\linewidth]{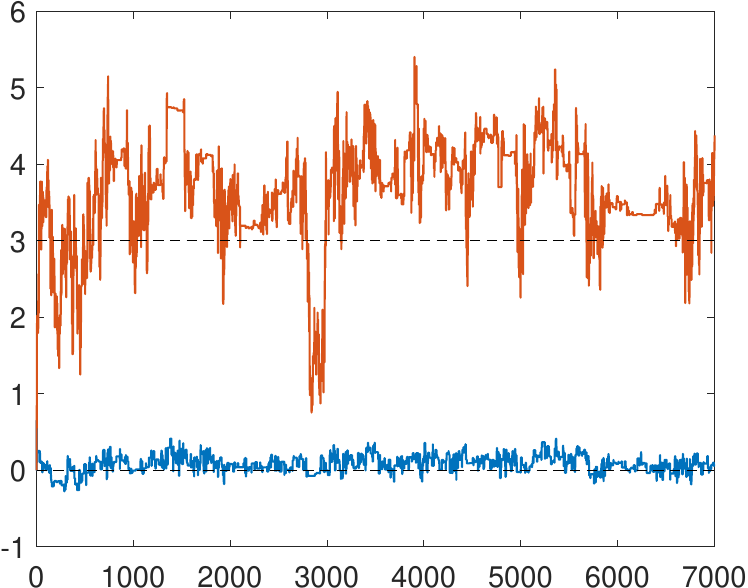}%
	\includegraphics[width=0.33\linewidth]{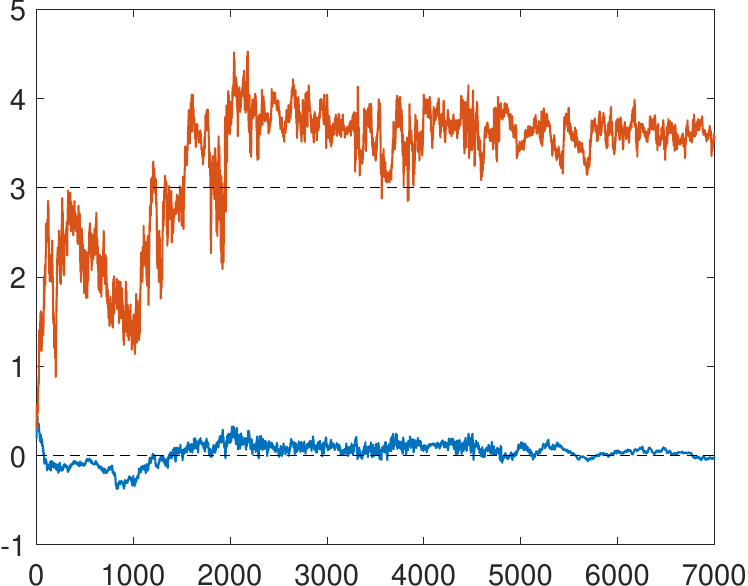}%
	\caption{Traces for parameters $\mu_1$ (blue) and $\mu_2$ (orange) for
			%standard HMC (left),
			\pmhmc (left), pseudo-marginal slice sampling (mid),
			and Particle Gibbs sampling (right), all with $N=128$ .}
			\label{fig:glmm_mu}
			\end{figure}

We ran \pmhmc, pseudo-marginal slice sampling \citep{MurrayG:2015}, and Particle Gibbs \citep{andrieudoucetholenstein2010} for 7\,000 iterations, all using $N=128$. % and a simple (naive) choice of importance distribution for the latent variables: $q(b_i) = \N(b_i \mid 0, 3^2)$. We also ran a Gibbs sampler
We note that Gibbs sampling type algorithms (of which our Particle Gibbs sampler is an example) are the \emph{de facto} standard methods for Bayesian GLMMs.
For the pseudo-marginal slice sampling we used a SS+MH algorithm where for $\theta$ we have one MH step for each component and for for $\uvar$ we used elliptical slice sampling.
% (Further details on the implementation of the different methods is given in the supplementary material.)
Figure~\ref{fig:glmm_mu} shows traces for the parameters $\mu_1$ and $\mu_2$,
with additional results reported in the \suppmat.
The \pmhmc method clearly outperforms the competing methods in terms of mixing (the computational cost per iteration is about 3.5 times higher for \pmhmc than for pseudo-marginal slice sampling in our implementation). In particular, compared to the results of Section~\ref{sec:sinc2}, we note that \pmhmc handles this more challenging model with a 13-dimensional $\theta$ much better than the pseudo-marginal slice sampling which appears here to struggle for high-dimensional~$\theta$.

However, we also note that  the pseudo-marginal HMC algorithm gets stuck for quite many iterations around iteration 1\thinspace000 (see Figure~\ref{fig:glmm_mu}).  Experiments suggest that this stickiness is an issue inherited from the marginal HMC (which pseudo-marginal HMC approximates) and is not specific to the pseudo-marginal HMC algorithm. Specifically, we have experienced that the pseudo-marginal HMC sampler tends to get stuck at large values of $\lambda_1$ or $\lambda_2$, \ie, in the right tail of the posterior for one of these parameters. A potential solution to address this issue is to use a state-dependent mass matrix as in the Riemannian manifold HMC \citep{GirolamiC:2011}.

			\section{Discussion\label{sec:discussion}}
			HMC methods cannot be implemented in scenarios where the likelihood function is intractable. However, we have shown here that if we have access to a non-negative unbiased likelihood estimator parameterized by normal random variables then it is possible to derive an algorithm which mimics the HMC algorithm having access to the exact likelihood. The resulting pseudo-marginal HMC algorithm replaces the original intractable gradient of the log-likelihood by the gradient of the log-likelihood estimator while preserving the target distribution as invariant distribution. Empirically we have observed that this algorithm can work significantly better than the pseudo-marginal MH algorithm as well as the standard HMC method sampling on the joint space for a fixed computational budget.

			However, whereas clear guidelines are available for optimizing the performance of the pseudo-marginal MH algorithm \citep{doucet2015efficient,Sherlock2015efficiency}, it is unclear how to address this problem for the pseudo-marginal HMC method.
			As we increase the number of samples $N$, the pseudo-marginal HMC algorithm can be seen as moving from an HMC sampling on the joint space (using a non-centered parameterization) to an HMC sampling on the marginal space. Thus, even the case $N=1$ results in a method which, in some cases, works well even for large $T$---we do not experience the same ``break down'' of the method as for pseudo-marginal MH when using a too small $N$ relative to $T$. In practice the number of samples $N$ is therefore a tuning parameter that needs to be chosen based on the geometry of the target posterior and traded off with the computational cost of the method.

			Finally, we have restricted ourselves here for presentation brevity to the pseudo-marginal approximation of a standard HMC algorithm using a constant mass matrix and a Verlet scheme. However, it is clear that the same ideas can be straightforwardly extended to more sophisticated HMC schemes such as the Riemannian manifold HMC  \citep{GirolamiC:2011} or schemes discretizing the associated Nos\'e-Hoover dynamics described in the \suppmat.

%\section*{Acknowledgments}
%Arnaud Doucet's research is supported by the UK Engineering and Physical Sciences Research Council, grants EP/K000276/1 and EP/K009850/1. Fredrik Lindsten's research is supported by the Swedish Research Council (VR) via the project \emph{Learning of complex dynamical systems} (Contract number: 637-2014-466).

%%\section*{References}

\bibliographystyle{abbrvnat}
\bibliography{ref}

\newpage
\appendix
\input{appendices_jmlr}

\end{document}

%% file: defs_jmlr.tex
\newcommand{\suppmat}{appendix\xspace}
\newcommand\range[2]{{#1\!:\!#2}}
\newcommand{\N}{\mathcal{N}}
\DeclareMathOperator\sinc{sinc}
\newcommand{\T}{T} % Number of latent variables // hard coded as T in most of the document!
\newcommand{\thsnd}[1]{#1\thinspace000}
\newcommand\pmhmc{PM-HMC\xspace}
\newcommand\hmc{HMC\xspace}
\newcommand{\reals}{\mathbb{R}}
\newcommand{\+}{\mathsf{T}} % Transpose

\newcommand{\convd}{\overset{d}{\longrightarrow}}
\newcommand{\convp}{\overset{p}{\longrightarrow}}

\newcommand{\E}{\mathbb{E}}
\newcommand{\eqdef}{\vcentcolon=}
\newcommand{\eqd}{\overset{d}{=}}

\newcommand{\lip}{\mathcal{L}}

\newcommand{\norm}[1]{\|#1\|}
\newcommand{\nset}{\mathbb{N}}

\newcommand{\parvec}[1]{\Theta[{#1}]}
\newcommand{\parvecest}[1]{\hat{\Theta}[{#1}]}

\newcommand{\parmatsmall}[1]{\begin{pmatrix}
	\thex{#1} \\
	\rhex{#1}
\end{pmatrix}}
\newcommand{\parmatsmallest}[1]{\begin{pmatrix}
	\thest{#1} \\
	\rhest{#1}
\end{pmatrix}}

\newcommand{\pvar}{{\bf{p}}}
\newcommand{\ph}{\hat{\pvar}}
\newcommand{\pest}[1]{\ph[#1]}
\newcommand{\pex}[1]{\pvar[#1]}
\newcommand{\pexp}[1]{\pvar(#1)}

\newcommand{\rh}{\hat{\rho}}
\newcommand{\rhest}[1]{\hat{\rho}[#1]}
\newcommand{\rhex}[1]{\rho[#1]}
\newcommand{\rhexp}[1]{\rho(#1)}
\newcommand{\rset}{\reals}

\newcommand{\that}{\hat{\theta}}
\newcommand{\thest}[1]{\hat{\theta}[#1]}
\newcommand{\thex}[1]{\theta[#1]}
\newcommand{\thexp}[1]{\theta(#1)}

\newcommand{\uvar}{{\bf{u}}}
\newcommand{\uh}{\hat{\uvar}}
\newcommand{\uest}[1]{\uh[#1]}
\newcommand{\uex}[1]{\uvar[#1]}
\newcommand{\uexp}[1]{\uvar(#1)}

\newcommand{\vvar}{{\bf{v}}}
\newcommand{\vh}{\hat{\vvar}}
\newcommand{\vest}[1]{\vh}

\newcommand{\wvar}{{\bf{w}}}
\newcommand{\wh}{\hat{\wvar}}
\newcommand{\west}[1]{\wh}
\newcommand{\westp}[1]{\wh'}
\newcommand{\wgtlow}{\underline{\omega}}
\newcommand{\wgtup}{\overline{\omega}}

%% file: appendices_jmlr.tex
%!TEX root = ./pmhmc_jmlr.tex

\section{The one step integrator}\label{app:int}
Taking one step of the integrator $\hat{\Phi}_{h} = \Phi^{A}_{h/2} \circ \Phi^{B}_{h} \circ \Phi^{A}_{h/2}$ (where $\Phi^{A}_h$ and $\Phi^{B}_{h}$ is given in~\eqref{eq:PhiA} and \eqref{eq:PhiB}) gives us the updating scheme from $\parvecest{\ell} = (\thest{\ell}, \rhest{\ell}, \uest{\ell}, \pest{\ell})$ to $\parvecest{\ell + 1}$ through the following equations,
\begin{align*}
	\thest{\ell + 1} &= \thest{\ell} + h \rhest{\ell} + \tfrac{h^2}{2} \nabla_{\theta}\left\{ \log p(\theta) + \log \widehat{p}(y \mid \theta, \pest{\ell}\sin(\tfrac{h}{2}) + \uest{\ell} \cos(\tfrac{h}{2})) \right\}_{\mid \theta = \thest{\ell} + \tfrac{h}{2}\rhest{\ell} },\\
	\rhest{\ell + 1} &= \rhest{\ell} + h \nabla_{\theta}\left\{ \log p(\theta) + \log \widehat{p}(y \mid \theta, \pest{\ell}\sin(\tfrac{h}{2}) + \uest{\ell} \cos(\tfrac{h}{2})) \right\}_{\mid \theta = \thest{\ell} + \tfrac{h}{2}\rhest{\ell} },\\
	\uest{\ell + 1} &= \pest{\ell} \sin(h) + \uest{\ell} \cos(h) + \sin(\tfrac{h}{2}) h  \nabla_{\uvar} \log \widehat{p}(y \mid \thest{\ell} + \tfrac{h}{2}\rhest{\ell} , \uvar)_{\mid \uvar = \pest{\ell}\sin(\tfrac{h}{2}) + \uest{\ell}\cos(\tfrac{h}{2})}, \\
	\pest{\ell + 1} &= \pest{\ell} \cos(h) - \uest{\ell} \sin(h) + \cos(\tfrac{h}{2}) h  \nabla_{\uvar} \log \widehat{p}(y \mid \thest{\ell} + \tfrac{h}{2}\rhest{\ell} , \uvar)_{\mid \uvar = \pest{\ell}\sin(\tfrac{h}{2}) + \uest{\ell}\cos(\tfrac{h}{2})}.
\end{align*}
The steps in between are omitted but can easily be checked. Some use of trigonometric identities has to be used to reach the final expressions.

\section{Convergence of simulated trajectories} %Hamiltonian dynamics}

\begin{lemma}\label{lem:difference}
	Let $f, \hat{f} : \rset^{n} \to \rset^{n}$ be continuous functions and let $x, \hat{x} : \nset \to \rset^{N}$ be the solution to the following difference equation:
	\begin{align}
		x[i+1] - x[i] = h \cdot f(x[i]), \quad \hat{x}[i+1] - \hat{x}[i] = h \cdot \hat{f}(\hat{x}[i]), \label{eq:prop:diff}
		%x_{h(i+1)} - x_{h i} = h \cdot f(x_{h i}) \quad \hat{x}_{h (i+1)} - \hat{x}_{h i} = h \cdot \hat{f}(\hat{x}_{h i }). \label{eq:prop:diff}
	\end{align}
	both initialized using $x[0] = \hat{x}[0]$. If $f$ is Lipschitz with constant $\lip$ then, for any $\ell \in \nset$
	\begin{align}
		\norm{\hat{x}[\ell] - x[\ell]} \leq \sum_{i=0}^{\ell - 1} h(1 + h \lip)^{\ell - (i+1)} \norm{ \hat{f}(\hat{x}[i]) - f(\hat{x}[i]) } \label{eq:diff:error:bound}
	\end{align}
\end{lemma}

\begin{proof}
	Using \eqref{eq:prop:diff} the proof is straightforward,
	\begin{align*}
		\norm{\hat{x}[i+1] - x[i+1]} & = \norm{\hat{x}[i] + h \cdot \hat{f}(\hat{x}[i]) - x[i] - h \cdot f(x[i]) } \\
		& \leq \norm{ \hat{x}[i] - x[i] } + h \norm{\hat{f}(\hat{x}[i]) - f(x[i])} \\
		& = \norm{ \hat{x}[i] - x[i] } + h \norm{\hat{f}(\hat{x}[i]) - f(\hat{x}[i]) + f(\hat{x}[i]) - f(x[i])} \\
		& \leq \norm{ \hat{x}[i] - x[i]} + h \norm{f(\hat{x}[i]) - f(x[i])} + h \norm{\hat{f}(\hat{x}[i]) - f(\hat{x}[i])} \\
		& \leq \norm{ \hat{x}[i] - x[i]} + h\lip\norm{\hat{x}[i] - x[i]} + h \norm{\hat{f}(\hat{x}[i]) - f(\hat{x}[i])} \\
		& = (1 + h \lip)\norm{\hat{x}[i] - x[i]} + h \norm{\hat{f}(\hat{x}[i]) - f(\hat{x}[i])}.
	\end{align*}
	Equation~\eqref{eq:diff:error:bound} then follows by repeating the recursion.
\end{proof}

For the proof of Proposition~\ref{prop:difference} we have to compare the result of the ideal \hmc algorithm with our proposed \pmhmc algorithm. Since they live on different spaces we augment the ideal \hmc algorithm to incorporate the $\uvar$ and $\pvar$ part in such a way that the $(\theta,\rho)$-marginal exactly follows the ideal \hmc algorithm. We do this by introducing the following time-stepper that will replace $\Phi^B_t$:
\begin{align*}
	\tilde{\Phi}^{B}_t : \begin{cases}
		\theta(t)=\theta(0),\\
		\rho(t)=\rho(0)+t\nabla_{\theta}\left\{ \log\thinspace p\left(\theta\right)+\nabla_{\theta}\log\thinspace p(y\mid\theta)\right\} _{|\theta=\theta(0)},\\
		\mathbf{u}(t)=\mathbf{u}(0),\\
		\mathbf{p}(t)=\mathbf{p}(0)+t\nabla_{\mathbf{u}}\log\thinspace\widehat{p}(y\mid\theta(0),\mathbf{u})_{|\mathbf{u}=\mathbf{u}(0)}.
	\end{cases}
\end{align*}
\fl{Add short explanation of how it differs from $\Phi^B$} \ja{done}
This differs from $\Phi^{B}$ by using the gradient from the exact posterior when updating $\rho(t)$.
Further we introduce $\tilde{\Phi}_h = \Phi^A_{h/2} \circ \tilde{\Phi}^{B}_h \circ \Phi^{A}_{h/2}$. Using this splitting operator as numerical integrator we have that the marginal $(\theta,\rho)$ will coincide exactly with the ideal \hmc algorithm.

\begin{lemma}\label{lem:lip}
	Assume that Assumption 3 holds. Also assume that $\nabla_{\theta} \log p(\theta \mid y)$ is Lipschitz with constant $L_0$, then it follows that $\tilde{\Phi}_h$ is Lipschitz with constant $\lip < \infty$ which does not depend on $N$.
\end{lemma}

The proof of Lemma~\ref{lem:lip} is postponed to Appendix~\ref{sec:lip}. Now we have all of the results needed to prove the bound on the difference between the output of the \pmhmc and the \hmc algorithm.
\\

\begin{proof}[Proof of Proposition~\ref{prop:difference}]
	Under the assumptions we have that Lemma~\ref{lem:lip} holds and thus that $\tilde{\Phi}_h$ is Lipschitz with constant $\lip < \infty$.

	As the space of the ideal HMC algorithm and the pseudo marginal HMC algorithm differ we cannot directly compare them. Thus we augment the space of the ideal HMC algorithm to reach the timestepper $\tilde{\Phi}_h = \Phi_h^A \circ \tilde{\Phi}^B_h \circ \Phi^A_h$ by adding the $\uvar{}$ and $\pvar{}$ part of the \pmhmc algorithm to the ideal \hmc algorithm. Notice that this is no longer a discretization of a Hamiltonian field but it leaves $(\theta, \rho)$ unchanged from the ideal \hmc algorithm. Let $\parvec{\ell} = (\thex{\ell}, \rhex{\ell}, \uex{\ell}, \pex{\ell}) = \tilde{\Phi}_{h\ell}(\parvec{0})$.
	\fl{I think it's worth writing out explicitly what $\parvec{\ell}$ is (in terms of the integrator).}\ja{added}
	We have that
	\begin{align*}
		\norm{\hat{\Phi}_{h}(\parvecest{\ell}) - \tilde{\Phi}_{h}(\parvecest{\ell})} = \sqrt{ \frac{h^4}{4} + h^2 }\norm{\nabla_{\theta} \log \left( \frac{\widehat{p}(y_{1:T} \mid \theta, \pest{\ell}\sin(\frac{h}{2}) + \uest{\ell} \cos(\frac{h}{2}))}{p(y_{1:T} \mid \theta)} \right)},
	\end{align*}
	by the assumption that $\tilde{\Phi}_h$ is Lipschitz with constant $\lip$ we have using Lemma~\ref{lem:difference} that
	\begin{align*}
		&\norm{\parmatsmallest{\ell} - \parmatsmall{\ell}} \leq \norm{\parvecest{\ell} - \parvec{\ell}} \\
		&\hspace{2 cm}\leq \sum_{i=0}^{\ell - 1} h (1 + h \lip)^{\ell - (i + 1)} \sqrt{ \frac{h^4}{4} + h^2 }\norm{\nabla_{\theta} \log \left( \frac{\widehat{p}(y_{1:T} \mid \theta, \pest{i}\sin(\frac{h}{2}) + \uest{i} \cos(\frac{h}{2}))}{p(y_{1:T} \mid \theta)} \right)}.
	\end{align*}
	%The proof is finalized by taking the expected value and noting that all the randomness is in the initial values.
	\fl{(Re)move this last sentence if we keep the current presentation of the Proposition (with the part where we take the expected value is outside of the proposition)} \ja{removed it}
\end{proof}

\section{Proof of CLT}\label{app:clt}

\begin{remark}\label{rem:mixture}
	For the proof of the CLT and the coming proof of the convergence of the acceptance probability the following ``trick'' will be used. Assume that $\uvar$ is drawn from the distribution $\bar{\pi}(\cdot \mid \theta)$. We use the fact that we can write this distribution as
\begin{align*}
	\bar{\pi}(\uvar \mid \theta) = \frac{1}{N}\sum_{j=1}^{N} \psi(\uvar_j \mid \theta) \prod_{i \neq j} \N(\uvar_i; 0_p, I_p),
\end{align*}
where $\psi(\, \cdot \mid \theta) \eqdef \N(\cdot \, ; 0_p, I_p) \times \varpi_{\theta}(y, \cdot)$. That is, a mixture distribution where we choose one component $j$ uniformly at random and sample that variable $\uvar_j$ from $\psi(\, \cdot \mid \theta)$ and the rest of the variables $\uvar_i : i\neq j$ from standard Gaussian distributions.

What can now be done, and will be done in the proofs below, is to introduce new variables $\west{} $ and $\westp{}_1$ such that $\west{} \sim \N(0_D,I_D)$ and $\westp{}_1 \sim \psi(\, \cdot \mid \theta)$. This will be used in the proofs to compute sums over the random variables in the following way, for some function $f$ we have that
\begin{equation*}
	\sum_j f(\uvar_j) \eqd \sum_j f(\west{}_j) + f(\westp{}_1) - f(\west{}_1).
\end{equation*}
The convergence of the right hand side is then split to deal with the sum where all random variables are Gaussian and the difference between a $\psi(\, \cdot \mid \theta)$ and Gaussian random variable, which is usually much easier then working with the left hand side.

\fl{The last sentence needs some further explanation. It's only when we compute some sum over all variables that we can use this trick.
$\sum f(\uvar_j) =^d \sum f(\west{}_j) + f(\westp{}_1)-f(\west{}_1)$
} \ja{changed the wording.}
\end{remark}

\begin{proof}[Proof of Proposition~\ref{prop:clt}]
	We start by proving the result for $i=0$, first we assume that $\uest{0} \sim \N(0_D, I_D)$ and secondly we relax this assumption by assuming that $\uest{0} \sim \bar{\pi}( \cdot \mid \theta).$ For ease of notation in the proof we will assume that we only have one observation ($T=1$) the extension to many observations is immediate.

	We assume that $\uest{0} \sim \N(0_D,I_D)$ and let $\vest{0} \eqdef \pest{0} \sin(\tfrac{h}{2}) + \uest{0}\cos(\tfrac{h}{2})$ then $\vest{0}$ also follows a $\N(0_d,I_D)$ distribution, since by definition we have that $\pest{0} \sim \N(0_D,I_D)$. Now we write
	\begin{align*}
		\log \widehat{p}(y \mid \theta, \vest{0}) - \log p(y \mid \theta) = \log \left\{ 1 + \frac{\widehat{p}(y \mid \theta, \vest{0}) - p(y \mid \theta)}{p(y \mid \theta)}\right\} = \log \left\{ 1 + \frac{\varepsilon_{N}(y, \vest{0} ; \theta)}{\sqrt{N}} \right\},
	\end{align*}
	where
	\begin{align*}
		\varepsilon_{N}(y, \vest{0}; \theta) \eqdef \sqrt{N} \frac{\widehat{p}(y \mid \theta, \vest{0}) - p(y \mid \theta)}{p(y \mid \theta)} = \frac{1}{\sqrt{N}} \sum_{i=1}^{N}\{ \varpi_{\theta}(y, \vest{0}_i) - 1 \}.
	\end{align*}
	Taking the gradient with respect to $\theta$ we get that
	\begin{align*}
		\sqrt{N} \{ \nabla_{\theta} \log \widehat{p}(y \mid \theta, \vest{0}) - \nabla_{\theta} p(y \mid \theta) \} = \frac{\nabla_{\theta} \varepsilon_{N}(y, \vest{0}; \theta)}{1 + \varepsilon_{N}(y, \vest{0}; \theta)/\sqrt{N}}.
	\end{align*}
	By the definitions we have that
	\begin{align*}
		&\E[\varepsilon_{N}(y, \vest{0}; \theta)^2] = \E[(\tfrac{1}{\sqrt{N}} \sum_{i=1}^{N}\{\varpi_{\theta}(y, \vest{0}_i) - 1\})^2] \\
		= &\frac{1}{N} \sum_{i=1}^N \sum_{j=1}^N\{ \E[\varpi_{\theta}(y, \vest{0}_i) \varpi_{\theta}(y, \vest{0}_j)] - \E[\varpi_{\theta}(y, \vest{0}_i)] - \E[\varpi_{\theta}(y, \vest{0}_j)] + 1 \} \\
		= & \E[\varpi_{\theta}(y, \vest{0}_1)^2] - 1 < \infty,
	\end{align*}
	where this upper limit is by assumption. Under the assumption that $\uest{0} \sim \N(0_D, I_D)$ we have that $\varepsilon_N(y, \vest{0}; \theta)/\sqrt{N} \convp 0$ by Chebyshev's inequality. Also noting that
	\begin{align*}
		\E[\nabla_{\theta} \varepsilon_N(y, \vest{0}; \theta)] = \sqrt{N} \E[ \nabla_{\theta}\varpi_{\theta}(y, \vest{0}_1) ] \overset{(1)}{=} \sqrt{N} \nabla_{\theta} \E[\varpi_{\theta}(y, \vest{0}_1)] = 0,
	\end{align*}
	where $(1)$ holds under the assumptions of the existence of the function $g(\cdot)$ such that $|\nabla_{\theta}\varpi_{\theta}(y,\uvar)| < g(\uvar)$ and $\E_{\N}[g(\uvar)] < \infty$ which then allows us to interchange the differential and integral. By the continuous mapping theorem, Slutsky's lemma and the standard CLT applied to $\nabla_{\theta}\varepsilon_N(y, \vest{0}; \theta)$ we get the desired results.

	So far we have only proven the result for the initialization of the algorithm, that is $\uest{0} \sim \N(0_D,I_D)$. When running the algorithm this will only be true for the very first iteration, after that the $\uh$ variables will either be accepted or rejected. In stationarity we will have that the pair $(\theta, \uh) \sim \bar{\pi}(\theta, \uh)$. Now we wish to prove the results under the assumption of having reached this distribution, that is $\uest{0} \sim \bar{\pi}( \cdot \mid \theta)$. We make use of Remark~\ref{rem:mixture} and introduce the variables $\west{0} \sim N(0,I_D)$ and $\westp{0}_1 \sim \psi(\cdot \mid \theta)$.

	% We use the fact that we can write this distribution as
	% \begin{align*}
	% 	\bar{\pi}(\uest{0} \mid \theta) = \frac{1}{N}\sum_{j=1}^{N} \psi(\uest{0}_j \mid \theta) \prod_{i \neq j} \N(\uest{0}_i; 0_p, I_p),
	% \end{align*}
	% where $\psi(\cdot \mid \theta) \eqdef \N(\cdot ; 0_p, I_p) \times \varpi_{\theta}(y, \cdot).$ This is a mixture distribution where we choose one coordinate uniformly at random and sample that coordinate from $\psi(\cdot \mid \theta)$ and the rest of the coordinates from standard Gaussian distributions.

	% Now introduce a set of variables $\west{0} \sim N(0,I_D)$ and $\westp{0}_1 \sim \psi(\cdot \mid \theta)$, then by adding and subtracting the term associated with $\pest{0}_1 \sin(\tfrac{h}{2}) + \west{0}_1 \cos(\tfrac{h}{2})$, for ease of notation we let $\pest{0}^{\wh,h} \eqdef \pest{0}\sin(\tfrac{h}{2}) + \west{0} \cos(\tfrac{h}{2})$ and write
	By adding and subtracting the term associated with $\pest{0}_1 \sin(\tfrac{h}{2}) + \west{0}_1 \cos(\tfrac{h}{2})$, for ease of notation we let $\pest{0}^{\wh,h} \eqdef \pest{0}\sin(\tfrac{h}{2}) + \west{0} \cos(\tfrac{h}{2})$ and write
	\begin{align*}
		&\sqrt{N} \{ \nabla_{\theta}\log\widehat{p}(y \mid \theta, \pest{0}\sin(\tfrac{h}{2}) + \uest{0} \cos(\tfrac{h}{2})) - \nabla_{\theta} \log p(y \mid \theta ) \} \\
		& \hspace{1 cm}= \frac{\nabla_{\theta}\varepsilon_{N}(y, \pest{0}\sin(\tfrac{h}{2}) + \uest{0}\cos(\tfrac{h}{2}))}{1 + \varepsilon_{N}(y, \pest{0}\sin(\tfrac{h}{2}) + \uest{0}\cos(\tfrac{h}{2}))/\sqrt{N}} \\
		& \hspace{1 cm} \eqd \frac{\nabla_{\theta}\varepsilon_{N}(y, \pest{0}^{\wh,h}; \theta) + \tfrac{1}{\sqrt{N}}(\nabla_{\theta}\varpi_{\theta}(y, \pest{0}_1 \sin(\tfrac{h}{2}) + \westp{0}_1 \cos(\tfrac{h}{2})) - \nabla_{\theta}\varpi_{\theta}(\pest{0}^{\wh,h}_1))}{1 + \tfrac{1}{\sqrt{N}}\varepsilon_{N}(y, \pest{0}^{\wh,h}; \theta) + \tfrac{1}{N}(\varpi_{\theta}(y, \pest{0}_1\sin(\tfrac{h}{2}) + \westp{0}_1 \cos(\tfrac{h}{2})) - \varpi_{\theta}(y, \pest{0}^{\wh,h}_1))},
	\end{align*}
	as previously we have that
	\begin{align*}
		\frac{1}{\sqrt{N}} \nabla_{\theta}\varpi_{\theta}(y, \pest{0}_1 \sin(\tfrac{h}{2}) + \westp{0}_1\cos(\tfrac{h}{2})) \convp 0& \quad \text{as } N \to \infty, \\
		\frac{1}{N}\varpi_{\theta}(y, \pest{0}_1 \sin(\tfrac{h}{2}) + \westp{0}_1\cos(\tfrac{h}{2})) \convp 0 &\quad \text{as } N \to \infty,
	\end{align*}
	with the same results holding when we use $\west{0}_1$ instead of $\westp{0}_1$. By the continuous mapping theorem and Slutsky's lemma we have the result for the initial step.

	Assume now that the following results hold at iteration $\ell$ for any $h > 0$,
	\begin{align*}
		\frac{1}{\sqrt{N}}\sum_{i=1}^{N} \nabla_{\theta}\varpi_{\theta}(y, \pest{\ell}_i \sin(h) + \uest{\ell}_i \cos(h)) &\convd \N(0, \sigma^2(\theta,y)), \quad &\text{as } N \to \infty, \\
		\frac{1}{N}\sum_{i=1}^{N}\varpi_{\theta}(y, \pest{\ell}_i \sin(h) + \uest{\ell}_i \cos(h)) &\convp 1, \quad &\text{as } N \to \infty.
	\end{align*}
	This gives us using Slutsky's lemma that
	\begin{align*}
		\sqrt{N}\nabla_{\theta}\log \left( \frac{\widehat{p}(y \mid \theta, \pest{\ell}\sin(h) + \uest{\ell}\cos(h))}{p(y \mid \theta)} \right) \convd \N(0, \sigma^2(\theta, y)), \quad \text{as } N \to \infty.
	\end{align*}
	Taking one step to iteration $\ell + 1$ we get that
	\begin{align*}
		&\pest{\ell + 1}\sin(\tfrac{h}{2}) + \uest{\ell+1}\cos(\tfrac{h}{2}) \\
		&\hspace{2cm}= \pest{\ell}\sin(\tfrac{3h}{2}) + \uest{\ell}\cos(\tfrac{3h}{2}) + \sin(h)h \nabla_{\uh}\log \widehat{p}(y \mid \theta, \uh) \mid_{\uh = \pest{\ell}\sin(\tfrac{h}{2}) + \uest{\ell}\cos(\tfrac{h}{2})},
	\end{align*}
	from the assumptions we have that $\varpi_{\theta}(y, \vvar) > 0$ and so we can use Lemma \ref{lem:clt} and Lemma \ref{lem:lln} which gives us that, using Slutsky's lemma
	\begin{align*}
		\sqrt{N} \nabla_{\theta}\log\left( \frac{\widehat{p}(y \mid \theta, \pest{\ell+1}\sin(\tfrac{h}{2}) + \uest{\ell+1}\cos(\tfrac{h}{2}))}{p(y \mid \theta)} \right) \convd \N(0, \sigma^2(\theta,y)), \quad \text{as } N \to \infty,
	\end{align*}
	where $\sigma^2(y,\theta) = \E[\{\nabla_{\theta}\varpi_{\theta}(y, \uvar)\}^{2}].$ This finishes the proof.

\end{proof}

\section{Proof of Lipschitz} \label{sec:lip}

\begin{proof}[Proof of Lemma~\ref{lem:lip}]
	By explicitly writing out $\tilde{\Phi}_h$ we get that
	\begin{align*}
		&\tilde{\Phi}_h(\Theta) = \tilde{\Phi}_h(\theta, \rho, \uvar, \pvar)  \\
		&= \begin{pmatrix}
			\theta + h \rho + \frac{h^2}{2} \nabla_{\theta}\{ \log p(\theta \mid y) \}_{\mid \theta = \theta + \frac{h}{2} \rho} \\
			\rho + h \nabla_{\theta}\{ \log p(\theta \mid y) \}_{\mid \theta = \theta + \frac{h}{2} \rho} \\
			\pvar \sin(h) + \uvar \cos(h) + \sin(\frac{h}{2}) h \nabla_{{\bf u}}\log \widehat{p}(y \mid \theta + \frac{h}{2} \rho, {\bf u})_{\mid {\bf u} = \pvar \sin(\frac{h}{2}) + \uvar\cos(\frac{h}{2})} \\
			\pvar \cos(h) - \uvar \sin(h) + \cos(\frac{h}{2}) h \nabla_{{\bf u}}\log \widehat{p}(y \mid \theta + \frac{h}{2} \rho, {\bf u})_{\mid {\bf u} = \pvar \sin(\frac{h}{2}) + \uvar\cos(\frac{h}{2})}
		\end{pmatrix},
	\end{align*}
	we need to find a $\lip$ such that
	\begin{align*}
		\norm{\tilde{\Phi}_{h}(\Theta) - \tilde{\Phi}_h(\hat{\Theta})} \leq \lip \norm{\Theta - \hat{\Theta}},
	\end{align*}
	or equivalently find $\lip^2$ such that
	\begin{align*}
		\norm{\tilde{\Phi}_{h}(\Theta) - \tilde{\Phi}_h(\hat{\Theta})}^2 \leq \lip^2\left\{\norm{\theta - \that}^2 + \norm{\rho - \rh}^2 +  \norm{\uvar - \uh}^2 + \norm{\pvar - \ph}^2 \right\}.
	\end{align*}
	We have that
	\begin{align*}
		&\norm{\tilde{\Phi}_h(\Theta) - \tilde{\Phi}_h(\hat{\Theta})}^2 \\
		\leq & \norm{\theta - \that}^2 + (1 + h^2)\norm{\rho - \rh}^2 + 2 \cos^2(h)\norm{\uvar - \uh} + 2 \sin^2(h)\norm{\pvar - \ph} \\
		+ &(h^2 + \tfrac{h^4}{4})\norm{\nabla_{\theta} \log p(\theta \mid y)_{\mid \theta = \theta + \tfrac{h}{2}\rho} - \nabla_{\theta} \log p(\theta \mid y)_{\mid \theta = \that + \tfrac{h}{2}\rh}}^2 \\
		+ & h^2 \norm{\nabla_{\uvar} \log \widehat{p}(y \mid \theta + \tfrac{h}{2}\rho, \uvar)_{\mid \uvar = \pvar \sin(\tfrac{h}{2}) + \uvar \cos(\tfrac{h}{2})} - \nabla_{\uvar} \log \widehat{p}(y \mid \that + \tfrac{h}{2}\rh, \uvar)_{\mid \uvar = \ph \sin(\tfrac{h}{2}) + \uh \cos(\tfrac{h}{2})}}^2.
	\end{align*}
	Under the assumption that $\nabla_{\theta} \log p(\theta \mid y)$ is Lipschitz with constant $L_0$ and assuming that $\nabla_{\uvar} \log \widehat{p}(y \mid \theta, \uvar)$ is Lipschitz with constant $L_1$ we have that
	\begin{align*}
		&\norm{\tilde{\Phi}_h(\Theta) - \tilde{\Phi}_h(\hat{\Theta})}^2 \\
		\leq & (1 + L_0^2(h^2 + \tfrac{h^4}{4}) + h^2 L_1^2)\norm{\theta - \that}^2 + (1 + h^2 + L_0^2 \tfrac{h^2}{4}(h^2 + \tfrac{h^4}{4}) + \tfrac{h^2}{4}L_1^2)\norm{\rho - \rh}^2 \\
		+ & (2\cos^2(h) + L_1^2 h^2 \cos^2(\tfrac{h}{2}))\norm{\uvar - \uh}^2 + (2 \sin^2(h) + L_1^2 h^2 \sin^2(\tfrac{h}{2}))\norm{\pvar - \ph}^2,
	\end{align*}
	it therefore holds that $\tilde{\Phi}_h$ is Lipschitz with constant $\lip = \sqrt{ 2 + h^2 + L_0^2(h^2 + \tfrac{h^4}{4}) + h^2 L_1^2}$. It remains to prove that $\nabla_{\uvar} \log \widehat{p}(y \mid \theta, \uvar)$ is Lipschitz with constant $L_1$ and that $L_1$ does not grow with $N$.

	For this we have that
	\begin{align} \nonumber
	&\left\Vert \nabla_{\mathbf{u}} \log\widehat{p}\left(  \left.  y\right\vert
	\that,\uh\right)  -\nabla_{\mathbf{u}}%
	\log\widehat{p}\left(  \left.  y\right\vert \theta,\mathbf{u}\right)
	\right\Vert \\ \label{eq:term1}
	&\quad\leq
	\left\Vert \nabla_{\mathbf{u}}\log\widehat{p}\left(  \left.  y\right\vert
	\theta, \uh\right)  -\nabla_{\mathbf{u}}%
	\log\widehat{p}\left(  \left.  y\right\vert \theta,\mathbf{u}\right)
	\right\Vert \\ \label{eq:term2}
	&\quad+ \left\Vert \nabla_{\mathbf{u}}\log\widehat{p}\left(  \left.  y\right\vert
	\that,\uh \right)  -\nabla_{\mathbf{u}}%
	\log\widehat{p}\left(  \left.  y\right\vert \theta,\uh\right)
	\right\Vert
\end{align}
Consider first term \eqref{eq:term1} (squared),
\begin{align*}
	&\left\Vert \nabla_{\mathbf{u}}\log\widehat{p}\left(  \left.  y\right\vert
	\theta,\uh\right)  -\nabla_{\mathbf{u}}%
	\log\widehat{p}\left(  \left.  y\right\vert \theta,\mathbf{u}\right)
	\right\Vert ^2 \\
	 &=
	 %TCIMACRO{\dsum \limits_{t=1}^{T}}%
	 %BeginExpansion
	 {\sum\limits_{t=1}^{T}}
	 %EndExpansion%
	 %TCIMACRO{\dsum _{i=1}^{N}}%
	 %BeginExpansion
	 {\sum_{i=1}^{N}}
	 %EndExpansion
	 \left\Vert \nabla_{\uvar_{t,i}}\log\widehat{p}\left(  \left.  y_{t}\right\vert
	 \theta,\uh_{t}\right)  -\nabla_{\uvar_{t,i}}\log\widehat{p}\left(
	 \left.  y_{t}\right\vert \theta,\mathbf{u}_{t}\right)  \right\Vert ^{2}
\end{align*}
We have
\[
\nabla_{\uvar_{t,i}}\log\widehat{p}\left(  \left.  y_{t}\right\vert \theta
,\mathbf{u}_{t}\right)  =\frac{\omega_{\theta}\left(  y_{t},\uvar_{t,i}\right)  }{%
%TCIMACRO{\dsum _{j=1}^{N}}%
%BeginExpansion
{\sum_{j=1}^{N}}
%EndExpansion
\omega_{\theta}\left(  y_{t},\uvar_{t,j}\right)  }\nabla_{\uvar_{t,i}}\log
\omega_{\theta}\left(  y_{t},\uvar_{t,i}\right)
\]
so
\begin{align}
\nonumber
& \left\Vert \nabla_{\uvar_{t,i}}\log\widehat{p}\left(  \left.  y_{t}\right\vert
\theta,\uh_{t}\right)  -\nabla_{\uvar_{t,i}}\log\widehat{p}\left(
\left.  y_{t}\right\vert \theta,\mathbf{u}_{t}\right)  \right\Vert ^{2}\\
\nonumber
& =\left\Vert \frac{\omega_{\theta}\left(  y_{t},\uh_{t,i}\right)  }{%
%TCIMACRO{\dsum _{j=1}^{N}}%
%BeginExpansion
{\sum_{j=1}^{N}}
%EndExpansion
\omega_{\theta}\left(  y_{t},\uh_{t,j}\right)  }\nabla_{\uvar_{t,i}}%
\log\omega_{\theta}\left(  y_{t},\uh_{t,i}\right)  -\frac
{\omega_{\theta}\left(  y_{t},\uvar_{t,i}\right)  }{%
%TCIMACRO{\dsum _{j=1}^{N}}%
%BeginExpansion
{\sum_{j=1}^{N}}
%EndExpansion
\omega_{\theta}\left(  y_{t},\uvar_{t,j}\right)  }\nabla_{\uvar_{t,i}}\log
\omega_{\theta}\left(  y_{t},\uvar_{t,i}\right)  \right\Vert ^{2}\\ \label{eq:term_a}
& \leq\left\{
\left\Vert \frac{\omega_{\theta}\left(  y_{t},\uh_{t,i}\right)
}{%
%TCIMACRO{\dsum _{j=1}^{N}}%
%BeginExpansion
{\sum_{j=1}^{N}}
%EndExpansion
\omega_{\theta}\left(  y_{t},\uh_{t,j}\right)  } \left( \nabla_{\uvar_{t,i}}%
\log\omega_{\theta}\left(  y_{t},\uh_{t,i}\right)  - \nabla_{\uvar_{t,i}}%
\log\omega_{\theta}\left(  y_{t},\uvar_{t,i}\right) \right) \right\Vert  \right. \\ \label{eq:term_b}
& \quad \left. +\left\Vert \left(  \frac{\omega_{\theta}\left(  y_{t},\uh_{t,i}\right)  }{%
%TCIMACRO{\dsum _{j=1}^{N}}%
%BeginExpansion
{\sum_{j=1}^{N}}
%EndExpansion
\omega_{\theta}\left(  y_{t},\uh_{t,j}\right)  }-\frac{\omega_{\theta
}\left(  y_{t},\uvar_{t,i}\right)  }{%
%TCIMACRO{\dsum _{j=1}^{N}}%
%BeginExpansion
{\sum_{j=1}^{N}}
%EndExpansion
\omega_{\theta}\left(  y_{t},\uvar_{t,j}\right)  }\right)  \nabla_{\uvar_{t,i}}%
\log\omega_{\theta}\left(  y_{t},\uvar_{t,i}\right)  \right\Vert \right\}^2
\end{align}
Under the Lipschitz assumption on the gradient of the log-weight-function the term on line \eqref{eq:term_a} is bounded by $M \Vert \uh_{t,i} - \uvar_{t,i}\Vert$.
 Furthermore, under the boundedness and Lipschitz assumptions on the weight function the term on line \eqref{eq:term_b} is bounded by
\begin{align*}
C& \left\vert \frac{\omega_{\theta}\left(  y_{t},\uh_{t,i}\right)  }{%
%TCIMACRO{\dsum _{j=1}^{N}}%
%BeginExpansion
{\sum_{j=1}^{N}}
%EndExpansion
\omega_{\theta}\left(  y_{t},\uh_{t,j}\right)  }-\frac{\omega_{\theta
}\left(  y_{t},\uvar_{t,i}\right)  }{%
%TCIMACRO{\dsum _{j=1}^{N}}%
%BeginExpansion
{\sum_{j=1}^{N}}
%EndExpansion
\omega_{\theta}\left(  y_{t},\uvar_{t,j}\right)  }\right\vert \\
& \leq C\left\vert \frac{\omega_{\theta}\left(  y_{t},\uh_{t,i}\right)
-\omega_{\theta}\left(  y_{t},\uvar_{t,i}\right)  }{%
%TCIMACRO{\dsum _{j=1}^{N}}%
%BeginExpansion
{\sum_{j=1}^{N}}
%EndExpansion
\omega_{\theta}\left(  y_{t},\uh_{t,j}\right)  }\right\vert
+ C  \frac{\omega_{\theta}\left(  y_{t},\uvar_{t,i}\right)  }{%
%TCIMACRO{\dsum _{j=1}^{N}}%
%BeginExpansion
{\sum_{j=1}^{N}}
%EndExpansion
\omega_{\theta}\left(  y_{t},\uvar_{t,j}\right)  }  \left\vert \frac{%
%TCIMACRO{\dsum _{j=1}^{N}}%
%BeginExpansion
{\sum_{j=1}^{N}}
%EndExpansion
\left\{  \omega_{\theta}\left(  y_{t},\uvar_{t,j}\right)  -\omega_{\theta}\left(
y_{t},\uh_{t,j}\right)  \right\}  }{%
%TCIMACRO{\dsum _{j=1}^{N}}%
%BeginExpansion
{\sum_{j=1}^{N}}
%EndExpansion
\omega_{\that}\left(  y_{t},\uh_{t,j}\right)  }\right\vert \\
& \leq\frac{C D}{N \underline{\omega}}\left\Vert
\uh_{t,i}-\uvar_{t,i}\right\Vert +\frac{CD}{N\underline{\omega}} \sum_{j=1}^{N}\left\Vert \uh_{t,j}-\uvar_{t,j}\right\Vert \\
& \leq \frac{2CD}{N \underline{\omega}}\sum_{j=1}^{N}\left\Vert \uh_{t,j}-\uvar_{t,j}\right\Vert %
\end{align*}
Put together we get for the term \eqref{eq:term1} (squared),
\begin{align*}
&\left\Vert \nabla_{\mathbf{u}}\log\widehat{p}\left(  \left.  y\right\vert
\that,\uh\right)  -\nabla_{\mathbf{u}}%
\log\widehat{p}\left(  \left.  y\right\vert \theta,\mathbf{u}\right)
\right\Vert ^2 \\
& \leq
%TCIMACRO{\dsum \limits_{t=1}^{T}}%
%BeginExpansion
{\sum\limits_{t=1}^{T}}
%EndExpansion%
%TCIMACRO{\dsum _{i=1}^{N}}%
%BeginExpansion
{\sum_{i=1}^{N}}
%EndExpansion
\left\{ M \Vert \uh_{t,i} - \uvar_{t,i}\Vert + \frac{2CD}{N \underline{\omega}}\sum_{j=1}^{N}\left\Vert \uh_{t,j}-\uvar_{t,j}\right\Vert  \right\}^2 \\
&=
%TCIMACRO{\dsum \limits_{t=1}^{T}}%
%BeginExpansion
{\sum\limits_{t=1}^{T}}
%EndExpansion%
%TCIMACRO{\dsum _{i=1}^{N}}%
%BeginExpansion
{\sum_{i=1}^{N}}
%EndExpansion
\left\{
 M^2 \Vert \uh_{t,i} - \uvar_{t,i}\Vert^2 + \left( \frac{2CD}{N\underline{\omega}} \sum_{j=1}^{N}\left\Vert \uh_{t,j}-\uvar_{t,j}\right\Vert \right)^2
+ \frac{4MCD}{N\underline{\omega}} \Vert \uh_{t,i} - \uvar_{t,i}\Vert  \sum_{j=1}^{N}\left\Vert \uh_{t,j}-\uvar_{t,j}\right\Vert \right\}
 \\
 &=
 %TCIMACRO{\dsum \limits_{t=1}^{T}}%
 %BeginExpansion
 {\sum\limits_{t=1}^{T}}
 %EndExpansion%
 \left\{
 M^2  \sum_{i=1}^{N} \Vert \uh_{t,i} - \uvar_{t,i}\Vert^2 + \left( \left[ \frac{2CD}{\underline{\omega}} \right]^2 + \frac{4MCD}{\underline{\omega}} \right) \times \frac{1}{N}\left( \sum_{j=1}^{N}\left\Vert \uh_{t,j}-\uvar_{t,j}\right\Vert \right)^2 \right\} \\
 &\leq
 %TCIMACRO{\dsum \limits_{t=1}^{T}}%
 %BeginExpansion
 {\sum\limits_{t=1}^{T}}
 %EndExpansion%
 \left\{
 M^2  \sum_{i=1}^{N} \Vert \uh_{t,i} - \uvar_{t,i}\Vert^2 + \left( \left[ \frac{2CD}{\underline{\omega}} \right]^2 + \frac{4MCD}{\underline{\omega}} \right) \times  \sum_{j=1}^{N}\left\Vert \uh_{t,j}-\uvar_{t,j}\right\Vert ^2 \right\} \\
 &= %\left( M^2 + \left[ \frac{2CD}{\underline{\omega}} \right]^2 + \frac{4MCD}{\underline{\omega}} \right) \Vert \uh - \mathbf{u} \Vert^2
 \left( M + \frac{2CD}{\underline{\omega}} \right)^2 \Vert \uh - \mathbf{u} \Vert^2
\end{align*}
where the inequality on the penultimate line follows from Jensen's inequality.

Next we address the term \eqref{eq:term2}. Analogously to above we have
\begin{align}
\nonumber
&\left\Vert \nabla_{\mathbf{u}}\log\widehat{p}\left(  \left.  y\right\vert
\theta^\prime,\mathbf{u}\right)  -\nabla_{\mathbf{u}}%
\log\widehat{p}\left(  \left.  y\right\vert \theta,\mathbf{u}\right)
\right\Vert ^2 \\ \label{eq:term2_expanded}
&=
%TCIMACRO{\dsum \limits_{t=1}^{T}}%
%BeginExpansion
{\sum\limits_{t=1}^{T}}
%EndExpansion%
%TCIMACRO{\dsum _{i=1}^{N}}%
%BeginExpansion
{\sum_{i=1}^{N}}
%EndExpansion
\left\Vert \nabla_{\uvar_{t,i}}\log\widehat{p}\left(  \left.  y_{t}\right\vert
\theta^\prime,\mathbf{u}_{t}\right)  -\nabla_{\uvar_{t,i}}\log\widehat{p}\left(
\left.  y_{t}\right\vert \theta,\mathbf{u}_{t}\right)  \right\Vert ^{2}
\end{align}
and
\begin{align*}
\nonumber
& \left\Vert \nabla_{\uvar_{t,i}}\log\widehat{p}\left(  \left.  y_{t}\right\vert
\theta^\prime,\mathbf{u}_{t}\right)  -\nabla_{\uvar_{t,i}}\log\widehat{p}\left(
\left.  y_{t}\right\vert \theta,\mathbf{u}_{t}\right)  \right\Vert ^{2}\\
\nonumber
& \leq\left\{
\left\Vert \frac{\omega_{\that}\left(  y_{t},\uvar_{t,i}\right)
}{%
%TCIMACRO{\dsum _{j=1}^{N}}%
%BeginExpansion
{\sum_{j=1}^{N}}
%EndExpansion
\omega_{\that}\left(  y_{t},\uvar_{t,j}\right)  } \left( \nabla_{\uvar_{t,i}}%
\log\omega_{\that}\left(  y_{t},\uvar_{t,i}\right)  - \nabla_{\uvar_{t,i}}%
\log\omega_{\theta}\left(  y_{t},\uvar_{t,i}\right) \right) \right\Vert  \right. \\
& \left. +\left\Vert \left(  \frac{\omega_{\that}\left(  y_{t},\uvar_{t,i}\right)  }{%
	%TCIMACRO{\dsum _{j=1}^{N}}%
	%BeginExpansion
	{\sum_{j=1}^{N}}
	%EndExpansion
	\omega_{\that}\left(  y_{t},\uvar_{t,j}\right)  }-\frac{\omega_{\theta
	}\left(  y_{t},\uvar_{t,i}\right)  }{%
	%TCIMACRO{\dsum _{j=1}^{N}}%
	%BeginExpansion
	{\sum_{j=1}^{N}}
	%EndExpansion
	\omega_{\theta}\left(  y_{t},\uvar_{t,j}\right)  }\right)  \nabla_{\uvar_{t,i}}%
\log\omega_{\theta}\left(  y_{t},\uvar_{t,i}\right)  \right\Vert \right\}^2 \\
& \leq\left\{
 \frac{\omega_{\that}\left(  y_{t},\uvar_{t,i}\right)
}{%
%TCIMACRO{\dsum _{j=1}^{N}}%
%BeginExpansion
{\sum_{j=1}^{N}}
%EndExpansion
\omega_{\that}\left(  y_{t},\uvar_{t,j}\right)  } M \Vert \that - \theta \Vert
+ C \frac{| \omega_{\that}\left(  y_{t},\uvar_{t,i}\right) - \omega_{\theta}\left(  y_{t},\uvar_{t,i}\right) |}{\sum_{j=1}^{N}
\omega_{\that}\left(  y_{t},\uvar_{t,j}\right)  }
\right. \\
& \quad + C \left.
\frac{\omega_{\theta}\left(  y_{t},\uvar_{t,i}\right)
}{%
%TCIMACRO{\dsum _{j=1}^{N}}%
%BeginExpansion
{\sum_{j=1}^{N}}
%EndExpansion
\omega_{\theta}\left(  y_{t},\uvar_{t,j}\right)
} \frac{ \left\vert  \sum_{j=1}^{N}
\omega_{\theta}\left(  y_{t},\uvar_{t,j}\right)-\omega_{\that}\left(  y_{t},\uvar_{t,j}\right) \right\vert }{\sum_{j=1}^{N}
\omega_{\that}\left(  y_{t},\uvar_{t,j}\right)}
 \right\}^2 \\
 & \leq\left\{
 \frac{\omega_{\that}\left(  y_{t},\uvar_{t,i}\right)
 }{%
 %TCIMACRO{\dsum _{j=1}^{N}}%
 %BeginExpansion
 {\sum_{j=1}^{N}}
 %EndExpansion
 \omega_{\that}\left(  y_{t},\uvar_{t,j}\right)  } M
+ \frac{CD}{\underline{\omega}} \left( \frac{1}{N}
 +
\frac{\omega_{\theta}\left(  y_{t},\uvar_{t,i}\right)
}{%
%TCIMACRO{\dsum _{j=1}^{N}}%
%BeginExpansion
{\sum_{j=1}^{N}}
%EndExpansion
\omega_{\theta}\left(  y_{t},\uvar_{t,j}\right)
} \right)
\right\}^2 \Vert \that - \theta \Vert ^2
\end{align*}
Plugging this expression into \eqref{eq:term2_expanded} we get
%expanding the square and summing over $i=1,\ldots,N$ we get
\begin{align*}
&\left\Vert \nabla_{\mathbf{u}}\log\widehat{p}\left(  \left.  y\right\vert
\theta^\prime,\mathbf{u}\right)  -\nabla_{\mathbf{u}}%
\log\widehat{p}\left(  \left.  y\right\vert \theta,\mathbf{u}\right)
\right\Vert ^2 \\
&\leq
%TCIMACRO{\dsum \limits_{t=1}^{T}}%
%BeginExpansion
{\sum\limits_{t=1}^{T}}
%EndExpansion%
%TCIMACRO{\dsum _{i=1}^{N}}%
%BeginExpansion
{\sum_{i=1}^{N}}
\left\{ \left(
\frac{\omega_{\that}\left(  y_{t},\uvar_{t,i}\right)
}{%
%TCIMACRO{\dsum _{j=1}^{N}}%
%BeginExpansion
{\sum_{j=1}^{N}}
%EndExpansion
\omega_{\that}\left(  y_{t},\uvar_{t,j}\right)  } \right)^2 M^2
+ \frac{C^2D^2}{\underline{\omega}^2} \left( \frac{1}{N}
+
\frac{\omega_{\theta}\left(  y_{t},\uvar_{t,i}\right)
}{%
%TCIMACRO{\dsum _{j=1}^{N}}%
%BeginExpansion
{\sum_{j=1}^{N}}
%EndExpansion
\omega_{\theta}\left(  y_{t},\uvar_{t,j}\right)
} \right)^2  \right. \\
&\quad + \left.  \frac{\omega_{\that}\left(  y_{t},\uvar_{t,i}\right)
}{%
%TCIMACRO{\dsum _{j=1}^{N}}%
%BeginExpansion
{\sum_{j=1}^{N}}
%EndExpansion
\omega_{\that}\left(  y_{t},\uvar_{t,j}\right)  }
\frac{2MCD}{\underline{\omega}} \left( \frac{1}{N}
+
\frac{\omega_{\theta}\left(  y_{t},\uvar_{t,i}\right)
}{%
%TCIMACRO{\dsum _{j=1}^{N}}%
%BeginExpansion
{\sum_{j=1}^{N}}
%EndExpansion
\omega_{\theta}\left(  y_{t},\uvar_{t,j}\right)
} \right)
\right\} \Vert \that - \theta \Vert ^2\\
&\leq
%TCIMACRO{\dsum \limits_{t=1}^{T}}%
%BeginExpansion
{\sum\limits_{t=1}^{T}}
%EndExpansion%
%TCIMACRO{\dsum _{i=1}^{N}}%
%BeginExpansion
{\sum_{i=1}^{N}}
\left\{
\frac{\omega_{\that}\left(  y_{t},\uvar_{t,i}\right)
}{%
%TCIMACRO{\dsum _{j=1}^{N}}%
%BeginExpansion
{\sum_{j=1}^{N}}
%EndExpansion
\omega_{\that}\left(  y_{t},\uvar_{t,j}\right)  }  M^2
+ \frac{C^2D^2}{\underline{\omega}^2} \left( \frac{1}{N^2}
+  \left(\frac{2}{N}+1\right)
\frac{\omega_{\theta}\left(  y_{t},\uvar_{t,i}\right)
}{%
%TCIMACRO{\dsum _{j=1}^{N}}%
%BeginExpansion
{\sum_{j=1}^{N}}
%EndExpansion
\omega_{\theta}\left(  y_{t},\uvar_{t,j}\right)
} \right)  \right. \\
&\quad + \left.  \frac{\omega_{\that}\left(  y_{t},\uvar_{t,i}\right)
}{%
%TCIMACRO{\dsum _{j=1}^{N}}%
%BeginExpansion
{\sum_{j=1}^{N}}
%EndExpansion
\omega_{\that}\left(  y_{t},\uvar_{t,j}\right)  }
\frac{2MCD}{\underline{\omega}} \left( \frac{1}{N}
+
1 \right)
\right\} \Vert \that - \theta \Vert ^2 \\
&=
%TCIMACRO{\dsum \limits_{t=1}^{T}}%
%BeginExpansion
{\sum\limits_{t=1}^{T}}
%EndExpansion%
%TCIMACRO{\dsum _{i=1}^{N}}%
%BeginExpansion
\left\{  M^2
+ \frac{C^2D^2}{\underline{\omega}^2} \left( \frac{1}{N}
+  \frac{2}{N}+1 \right)  +
\frac{2MCD}{\underline{\omega}} \left( \frac{1}{N}
+
1 \right)
\right\} \Vert \that - \theta \Vert ^2 \\
&\leq T\left(M+\frac{2CD}{\underline{\omega}}\right)^2  \Vert \that - \theta \Vert ^2
\end{align*}
It follows that
\begin{align*}
&\left\Vert \nabla_{\mathbf{u}}\log\widehat{p}\left(  \left.  y\right\vert
\that,\uh\right)  -\nabla_{\mathbf{u}}%
\log\widehat{p}\left(  \left.  y\right\vert \theta,\mathbf{u}\right)
\right\Vert \\
&\quad\leq
\left(M+\frac{2CD}{\underline{\omega}}\right)  \Vert \uh - \mathbf{u} \Vert +
\sqrt{T}\left(M+\frac{2CD}{\underline{\omega}}\right)  \Vert \that - \theta \Vert \\
&\quad\leq (\sqrt{T} + 1) \times \left(M+\frac{2CD}{\underline{\omega}}\right)  \left\Vert \begin{pmatrix}
\that \\ \uh
\end{pmatrix} - \begin{pmatrix}
\theta \\ \mathbf{u}
\end{pmatrix} \right\Vert.
\end{align*}
	
\end{proof}

\section{Proof of convergence of acceptance probability}

\begin{proof}[Proof of Proposition~\ref{prop:acc}]
	There are two parts of this proof, first we will show that the log-likelihood estimator $\log \widehat{p}(y \mid \theta, \uvar)$ converges to the log-likelihood $\log p(y \mid \theta)$ as $N \to \infty$. The second part of the proof is to show that the remaining $\uvar$ and $\pvar$ parts of the acceptance probability vanishes as $N$ increases.

	The first part follows by the proof of Proposition~\ref{prop:clt}, see Appendix~\ref{app:clt}. For completeness we repeat that part here, we do the proof by induction to show that for all $\ell$ we have that,
	\begin{align*}
		\log \widehat{p}(y \mid \theta, \uest{\ell}) \convp \log p(y \mid \theta), \quad \text{as } N \to \infty.
	\end{align*}
	It turns out that it is needed to show this result in a more general setting, that is for any $h > 0$ we wish to show that
	\begin{align}
		\widehat{p}(y \mid \theta, \pest{\ell}\sin(h) + \uest{\ell} \cos(h)) \convp p(y \mid \theta), \quad \text{as } N \to \infty. \label{eq:likelihood:convergence}
	\end{align}

	When $\ell = 0$ we prove the result in two different settings. First we assume that $\uest{0} \sim \N(0_D,I_D)$ and the result is clear since~\eqref{eq:likelihoodestimator} is just the likelihood of the importance sampling estimator under the assumption of Gaussian proposal distribution. Secondly we look at the case when $\uest{0} \sim \bar{\pi}(\, \cdot \mid \theta)$, we again introduce the variables $\westp{}_1 \sim \psi(\, \cdot \mid \theta)$ and $\west{} \sim \N(0_D,I_D)$ by the use of Remark~\ref{rem:mixture}, we get that
	\begin{align*}
		\widehat{p}(y \mid \theta, \pest{0}\sin(h) + \uest{0} \cos(h)) = \frac{1}{N}\sum_{i=1}^{N} \omega_{\theta}(\pest{0}_i \sin(h) + \uest{0}_i \cos(h)) \\
		\eqd \frac{1}{N} \omega_{\theta}(\pest{0}_1\sin(h) + \westp{}_1\cos(h)) + \frac{1}{N} \sum_{i=2}^{N} \omega_{\theta}(\pest{0}_i \sin(h) + \west{}_i \cos(h)),
	\end{align*}
	here the first part converges to zero and the second part converges to the likelihood. By Slutsky's lemma and the continuous mapping theorem we have the result.

	Assume now that~\eqref{eq:likelihood:convergence} holds for any $h > 0$. We then have for $\ell + 1$ and any $h'>0$ that might differ from the integration step that
	\begin{align*}
		&\widehat{p}(y \mid \theta, \pest{\ell + 1}\sin(h') + \uest{\ell + 1}\cos(h')) \\
		&= \widehat{p}\Big(y \mid \theta, \pest{\ell}\sin(h' + h) + \uest{\ell}\cos(h' + h) \\
		&\hspace{2 cm}+ h \sin(h' + \tfrac{h}{2}) \nabla_{\uvar} \log \widehat{p}(y \mid \theta, \uvar)_{\mid \uvar = \pest{\ell}\sin(\tfrac{h}{2}) + \uest{\ell}\cos(\tfrac{h}{2})}\Big) \\
		&= \frac{1}{N}\sum_{i=1}^{N}\omega_{\theta}\Big(y, \pest{\ell}\sin(h' + h) + \uest{\ell}\cos(h' + h) \\
		&\hspace{2 cm}+ h \sin(h' + \tfrac{h}{2}) \nabla_{\uvar} \log \widehat{p}(y \mid \theta, \uvar)_{\mid \uvar = \pest{\ell}\sin(\tfrac{h}{2}) + \uest{\ell}\cos(\tfrac{h}{2})} \Big),
	\end{align*}
	which converges to $p(y \mid \theta)$ by Lemma~\ref{lem:lln}. The result now follows by the continuous mapping theorem.
	% given by the prooFor the first part we will look at the convergence of $\widehat{p}(y \mid \theta, \uvar)$ and then use the continuous mapping theorem to get the desired result. We prove this by induction, we begin with the case when $\ell = 0$ and look at the two cases $\uest{0} \sim \N(0,I_D)$ and $\uest{0} \sim \bar{\pi}(\cdot \mid \theta)$.

	Let us take a look at the second part. For this part we will show that
	\begin{align*}
		\uest{0}^T \uest{0} + \pest{0}^T \pest{0} - \uest{L}^T \uest{L} - \pest{L}^T \pest{L} \convp 0, \quad \text{as } N \to \infty.
	\end{align*}
	We do this by rewriting this expression using a telescoping sum to
	\begin{align*}
		&\uest{0}^T \uest{0} + \pest{0}^T \pest{0} - \uest{L}^T \uest{L} - \pest{L}^T \pest{L} \\
		=& \sum_{i=1}^{N} \left( \uest{0}_{i}^2 + \pest{0}_{i}^{2} - \uest{L}_{i}^{2} - \pest{L}_{i}^{2} \right) \\
		=& \sum_{\ell = 0}^{L-1} \sum_{i=1}^{N} \left( \uest{\ell }_{i}^2 + \pest{\ell}_{i}^2 - \uest{\ell+1}_{i}^2 - \pest{\ell+1}_{i}^2 \right).
	\end{align*}
	What remains to prove is that, for all $\ell$,
	\begin{align*}
		\sum_{i=1}^{N} \left( \uest{\ell }_{i}^2 + \pest{\ell}_{i}^2 - \uest{\ell+1}_{i}^2 - \pest{\ell+1}_{i}^2 \right) \convp 0, \quad \text{as } N \to \infty.
	\end{align*}

	By taking one step of the integrator we have that, see Appendix~\ref{app:int}
	\begin{align*}
		\uest{\ell+1}_{i} &= \pest{\ell}_{i} \sin(h) + \uest{\ell }_{i} \cos(h) + \sin(\tfrac{h}{2}) h \nabla_{\uvar_i} \log \widehat{p}(y \mid \theta, \uvar_i)_{|\uvar_i = \pest{\ell}_i\sin(\tfrac{h}{2}) + \uest{\ell}_i\cos(\tfrac{h}{2})} \\
		\pest{\ell+1}_{i} &= \pest{\ell}_{i} \cos(h) - \uest{\ell}_{i} \sin(h) + \cos(\tfrac{h}{2}) h \nabla_{\uvar_i} \log \widehat{p}(y \mid \theta, \uvar_i)_{|\uvar_i = \pest{\ell}_i\sin(\tfrac{h}{2}) + \uest{\ell}_i\cos(\tfrac{h}{2})},
	\end{align*}
	combining these we get that, using some trigonometric equalities,
	\begin{align*}
		&\uest{\ell + 1}_{i}^2 + \pest{\ell + 1}_{i}^2 = \pest{\ell}_{i}^2 + \uest{\ell}_{i}^2 + h^2 (\nabla_{\uvar_i} \log \widehat{p}(y \mid \theta, \uvar_i)_{|\uvar_i = \pest{\ell}_i\sin(\tfrac{h}{2}) + \uest{\ell}_i\cos(\tfrac{h}{2})} )^2  \\
		&\hspace{2 cm}+ 2 h (\pest{\ell}_i \cos(\tfrac{h}{2}) + \uest{\ell}_i\sin(\tfrac{h}{2})) \nabla_{\uvar_i} \log \widehat{p}(y \mid \theta, \uvar_i)_{|\uvar_i = \pest{\ell}_i\sin(\tfrac{h}{2}) + \uest{\ell}_i\cos(\tfrac{h}{2})}.
	\end{align*}

	We now need to show two results, both holding as $N \to \infty$,
	\begin{align*}
		\text{(i)} \quad & \sum_{i=1}^{N} (\nabla_{\uvar_i} \log \widehat{p}(y \mid \theta, \uvar_i)_{|\uvar_i = \pest{\ell}_i\sin(\tfrac{h}{2}) + \uest{\ell}_i\cos(\tfrac{h}{2})} )^2  \convp 0, \\
		\text{(ii)} \quad & \sum_{i=1}^{N} (\pest{\ell}_i \cos(\tfrac{h}{2}) + \uest{\ell}_i\sin(\tfrac{h}{2})) \nabla_{\uvar_i} \log \widehat{p}(y \mid \theta, \uvar_i)_{|\uvar_i = \pest{\ell}_i\sin(\tfrac{h}{2}) + \uest{\ell}_i\cos(\tfrac{h}{2})} \convp 0.
	\end{align*}
	
	Starting with (i) we have that, by the assumptions on the weight functions, that
	\begin{align*}
		\sum_{i=1}^{N} (\nabla_{\uvar_i} \log \widehat{p}(y \mid \theta, \uvar_i)
		%_{|\uvar_i = \pest{\ell}_i\sin(\tfrac{h}{2}) + \uest{\ell}_i\cos(\tfrac{h}{2})}
		 )^2 \leq \sum_{i=1}^{N} \frac{\wgtup^2}{(\sum_{i=1}^N \wgtlow )^2} C^2 = \frac{\wgtup^2 C}{N \wgtlow^2} \to 0, \quad \text{as } N \to \infty.
	\end{align*}

	For (ii) using the same bounds related to the gradient we get
	\begin{align*}
		&\left| \sum_{i=1}^{N} (\pest{\ell}_i \cos(\tfrac{h}{2}) + \uest{\ell}_i\sin(\tfrac{h}{2})) \nabla_{\uvar_i} \log \widehat{p}(y \mid \theta, \uvar_i)
		%_{|\uvar_i = \pest{\ell}_i\sin(\tfrac{h}{2}) + \uest{\ell}_i\cos(\tfrac{h}{2})}
		 \right| \\
		& \hspace{5 cm}\leq \frac{1}{N} \frac{C \wgtup}{\wgtlow} \sum_{i}^{N} (\pest{\ell}_i \cos(\tfrac{h}{2}) + \uest{\ell}_i\sin(\tfrac{h}{2})).
	\end{align*}
	We now need to prove that
	\begin{align*}
		\frac{1}{N} \sum_{i}^{N} (\pest{\ell}_i \cos(\tfrac{h}{2}) + \uest{\ell}_i\sin(\tfrac{h}{2})) \convp 0, \quad \text{as } N \to \infty.
	\end{align*}
	We will prove this for every $\ell \geq 0$ and for any value of $h$ and this will be done using a proof of induction similar to what is done in the proof of the CLT in Appendix~\ref{app:clt}.

	We begin when $\ell = 0$, we have that $\pest{0} \sim \N(0_D,I_D)$ and assume first that $\uest{0} \sim \N(0,I_D)$. Then the result is trivial. If we instead assume that $\uest{0} \sim \bar{\pi}( \cdot \mid \theta)$. We again use Remark~\ref{rem:mixture} and introduce the set of variables $\west{}_{1} \sim \psi(\cdot \mid \theta)$ and $\west{}_{2:N} \sim \N(0_p,I_p)$. We then have that, for any $h > 0$,
	\begin{align*}
		&\frac{1}{N}\sum_{i=1}^{N} (\pest{0}_i \cos(h) + \uest{0}_i \sin(h)) \eqd \frac{1}{N}\sum_{i=1}^{N} (\pest{0}_i \cos(h) + \west{}_{i} \sin(h)) \\
		&= \frac{1}{N} (\pest{0}_1 \cos(h) + \west{}_1 \sin(h)) + \frac{1}{N}\sum_{i=2}^{N} (\pest{0}_i \cos(h) + \west{}_i \sin(h)).
	\end{align*}
	Both of these terms now converge to 0, the first part trivially does this, while the second part is a sum of standard Gaussian variables and by the law of large numbers this sum converges to 0.

	Assume now that for any $h > 0$ we have that
	\begin{align*}
		\frac{1}{N}\sum_{i=1}^{N} (\pest{\ell}_i \cos(h) + \uest{\ell}_i \sin(h)) \convp 0, \quad \text{as } N \to \infty.
	\end{align*}
	We now look at $\ell + 1$, by taking one step of the numerical integrator we get that, for any $h' > 0$ which may be different then the integration step length,
	\begin{align*}
		&\frac{1}{N}\sum_{i=1}^{N} (\pest{\ell + 1}_i \cos(h') + \uest{\ell + 1}_{i} \sin(h')) \\
		=& \frac{1}{N} \sum_{i=1}^{N}(\pest{\ell}_i\sin(h + h') + \uest{\ell}_i\cos(h + h') \\
		&\hspace{2 cm} + h \sin(\tfrac{h}{2} + h')\nabla_{\uvar_i}\log \widehat{p}(y \mid \theta, \uvar_i)_{\mid \uvar_i = \pest{\ell}_i\sin(\tfrac{h}{2}) + \uest{\ell}_i\cos(\tfrac{h}{2})}) \\
		=& \frac{1}{N}\sum_{i=1}^{N}(\pest{\ell}_{i} \sin(h + h') + \uest{\ell}_{i} \cos(h + h')) \\
		& \hspace{2 cm}+ \frac{1}{N}\sum_{i=1}^{N}h \sin(\tfrac{h}{2} + h')\nabla_{\uvar_i}\log \widehat{p}(y \mid \theta, \uvar_i)_{\mid \uvar_i = \pest{\ell}_i\sin(\tfrac{h}{2}) + \uest{\ell}_i\cos(\tfrac{h}{2})},
	\end{align*}
	where the first term converges to 0 in probability by the induction hypothesis and the second sum converges to 0 in probability in the same way as (i) above. Thus completing the proof.
\end{proof}

\section{Extension to Nos\'e-Hoover dynamics}

The Hamiltonian dynamics presented in (\ref{eq:HMCdyn}),
respectively (\ref{eq:pseudoHMCflow}),
preserves the Hamiltonian (\ref{eq:Hamiltonianmarginal}), respectively
(\ref{eq:extendedHamiltonian}). As mentioned earlier, it is thus
necessary to randomize periodically the momentum to explore the target
distribution of interest. On the contrary, Nos\'e-Hoover type dynamics
do not preserve the Hamiltonian but keep the target distribution of
interest invariant \cite[Chapter 8]{leimkhulermatthews2015}. However,
they are not necessarily ergodic but can perform well and
it is similarly possible to randomize them to ensure ergodicity.

Compared to the Hamiltonian dynamics (\ref{eq:HMCdyn}),
the Nos\'e-Hoover dynamics is given by
\begin{align}
\frac{\mathrm{d}\theta}{\mathrm{d}t} & =\rho,\label{eq:exactNoseHoover1}\\
\frac{\mathrm{d}\rho}{\mathrm{d}t} & =\nabla_{\theta}\log p\left(\theta\right)+\nabla_{\theta}\log p(y\mid\theta)-\xi\rho,\label{eq:exactNoseHoover2}\\
\frac{\mathrm{d}\xi}{\mathrm{d}t} & =\mu^{-1}\left(\rho^{T}\rho-d\right)\label{eq:exactNoseHoover3}
\end{align}
where $\xi\in\mathbb{R}$ is the so-called thermostat. It is easy
to check that this flow preserves
\[
\pi(\theta,\rho,\xi)=\pi(\theta,\rho)\mathcal{N}\left(\xi;0,\mu^{-1}\right)
\]
invariant; i.e. if $\left(\theta(0),\rho(0),\xi(0)\right)\sim\pi$
then $\left(\theta(t),\rho(t),\xi(t)\right)\sim\pi$
for any $t>0$. This dynamics can be straightforwardly extended to
the intractable likelihood case to obtain
\begin{align}
\frac{\mathrm{d}\theta}{\mathrm{d}t} & =\rho,\label{eq:extendedNoseHoover1}\\
\frac{\mathrm{d}\rho}{\mathrm{d}t} & =\nabla_{\theta}\thinspace\log p\left(\theta\right)+\nabla_{\theta}\log\thinspace\widehat{p}(y\mid\theta,\mathbf{u})-\xi\rho,\label{eq:extendedNoseHoover2}\\
\frac{\mathrm{d}\mathbf{u}}{\mathrm{d}t} & =\mathbf{p},\label{eq:extendedNoseHoover3}\\
\frac{\mathrm{d}\mathbf{p}}{\mathrm{d}t} & =-\mathbf{u}+\nabla_{\mathbf{u}}\log\thinspace\widehat{p}(y\mid\theta,\mathbf{u})-\xi\mathbf{p},\label{eq:extendedNoseHoover4}\\
\frac{\mathrm{d}\xi}{\mathrm{d}t} & =\mu^{-1}\left(\rho^{T}\rho+\mathbf{p}^{T}\mathbf{p}-\left(d+D\right)\right).\label{eq:extendedNoseHoover5}
\end{align}
This flow preserves
\[
\overline{\pi}(\theta,\rho,\mathbf{u},\mathbf{p},\xi)=\overline{\pi}(\theta,\rho,\mathbf{u},\mathbf{p})\mathcal{N}\left(\xi;0,\mu^{-1}\right).
\]
It would also be possible to use the thermostat $\xi$ to only regulate
$\rho$ as in (\ref{eq:exactNoseHoover1})-(\ref{eq:exactNoseHoover3}).

It is possible to combine Nos\'e-Hoover numerical integrators with an
MH\ accept-reject step to preserve the invariant distribution \cite{leimkhulerreich2009}.
A\ weakness of this approach is that the MH accept-reject step is
not only dependent of the difference between the ratio of the target
at the proposed state and the target at the initial state but involves
an additional Jacobian factor.
We will not explore further these approaches here, which will be the
subject of future work.

\section{Details about the numerical illustrations and additional results}
\subsection{Diffraction model}
A normal $\N(0,10^2)$ prior was used for each component of $\theta$.
We used $h=0.02$ and $L=50$ in all cases for simplicity, resulting in average acceptance probabilities in the range 0.6--0.8.

\begin{figure}[H]
	\centering
	\includegraphics[width=0.5\linewidth]{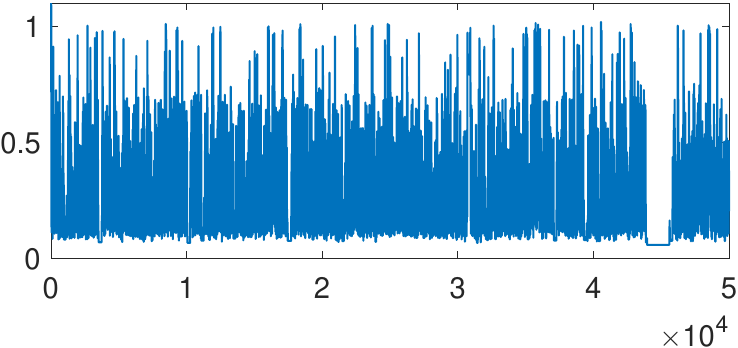}%
	\includegraphics[width=0.5\linewidth]{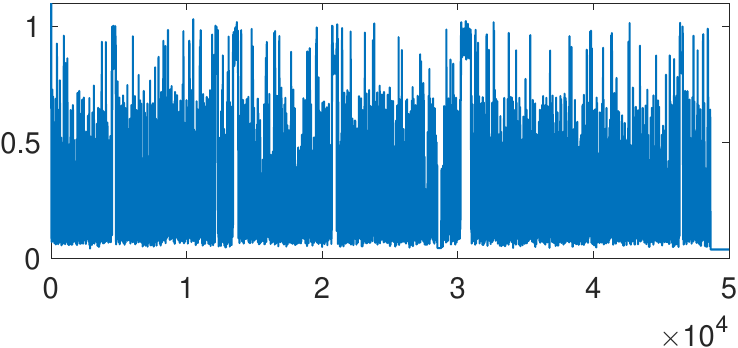}\\
	\includegraphics[width=0.5\linewidth]{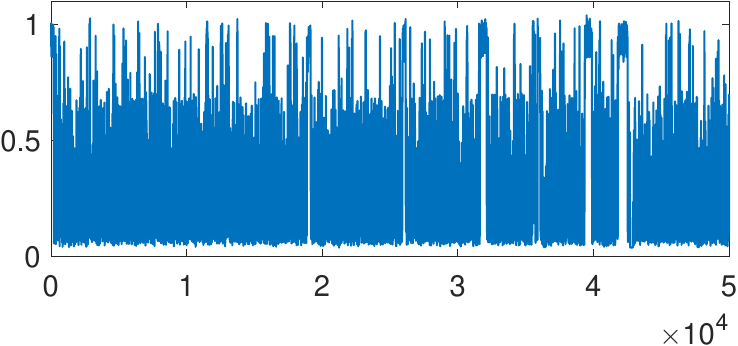}%
	\includegraphics[width=0.5\linewidth]{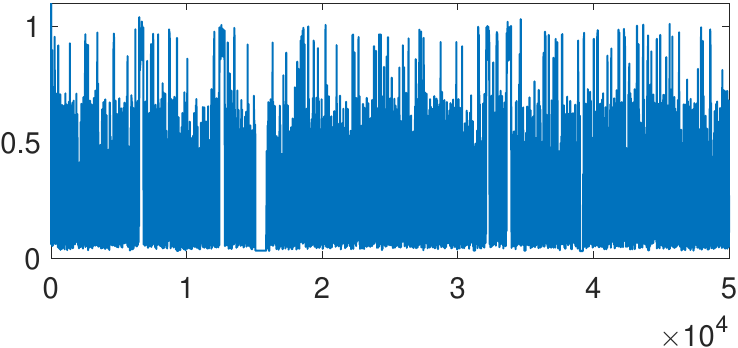}\vspace{-1.5\baselineskip}
	\caption{Traces for parameter $\lambda$ for the \textbf{PM-HMC sampler}.From top left to bottom right: $N=16$, $N=64$, $N=128$, $N=256$.}
	\label{fig:sinc2_pmhmc}
\end{figure}

\begin{figure}[H]
	\centering
	\includegraphics[width=0.5\linewidth]{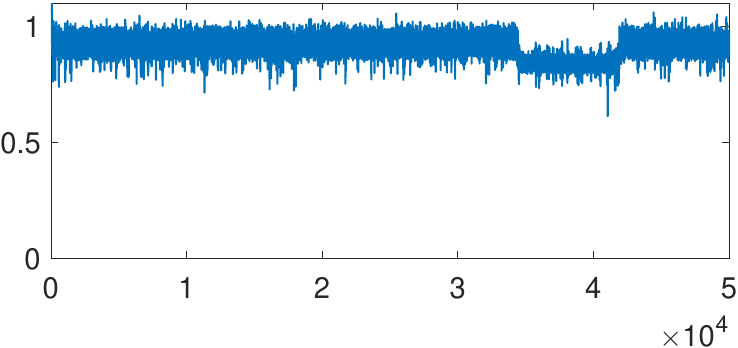}%\vspace{-1.5\baselineskip}%
	\caption{Traces for parameter $\lambda$ for the \textbf{standard HMC sampler}, using a non-centered parameterisation. (Note that the sampler completely misses one mode of the posterior.)}
	\label{fig:sinc2_hmc}
\end{figure}

\begin{figure}[H]
	\centering
	\includegraphics[width=0.5\linewidth]{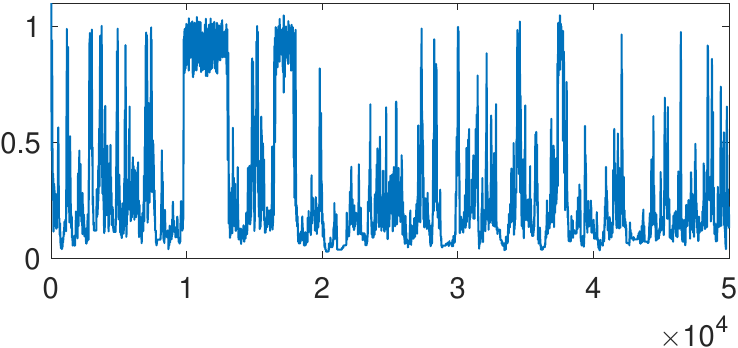}%
	\includegraphics[width=0.5\linewidth]{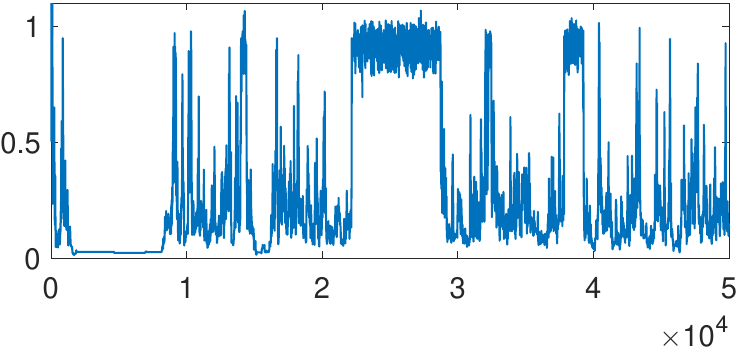}
	\vspace{-1.5\baselineskip}
	\caption{Traces for parameter $\lambda$ for the \textbf{pseudo-marginal MH sampler} with $N$ particles. Left, $N=512$. Right, $N=1024$.}
	\label{fig:sinc2_pmmh}
\end{figure}

\begin{figure}[H]
	\centering
	\includegraphics[width=0.5\linewidth]{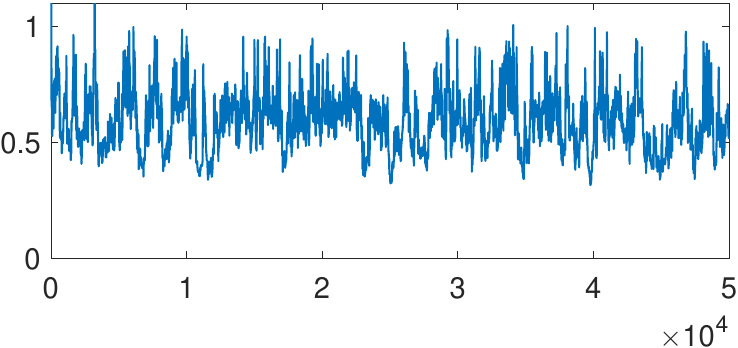}%
	\includegraphics[width=0.5\linewidth]{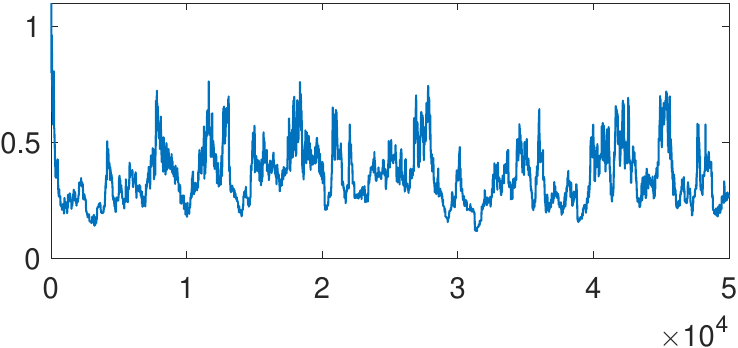}\\
	\includegraphics[width=0.5\linewidth]{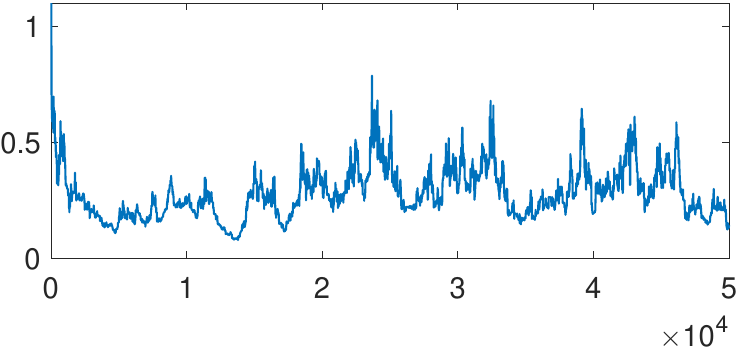}%
	\includegraphics[width=0.5\linewidth]{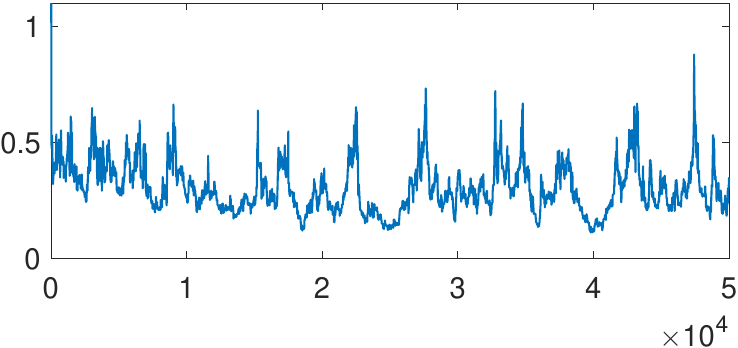}\vspace{-1.5\baselineskip}
	\caption{Traces for parameter $\lambda$ for the \textbf{Gibbs sampler} using conditional importance sampling kernels with $N$ particles to update the latent variables. From top left to bottom right: $N=16$, $N=64$, $N=128$, $N=256$.}
	\label{fig:sinc2_gibbs}
\end{figure}

\begin{figure}[H]
	\centering
	\includegraphics[width=0.5\linewidth]{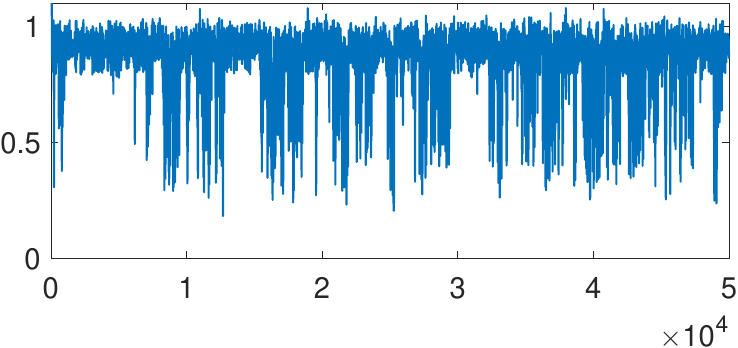}%
	\includegraphics[width=0.5\linewidth]{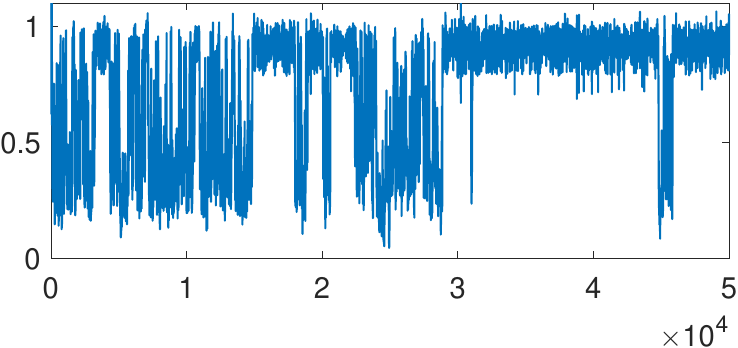}\\
	\includegraphics[width=0.5\linewidth]{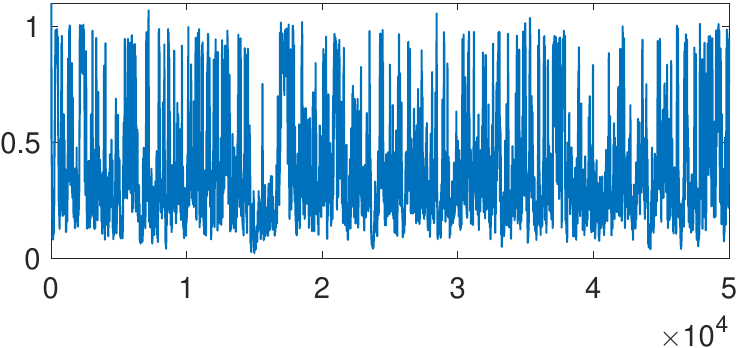}%
	\includegraphics[width=0.5\linewidth]{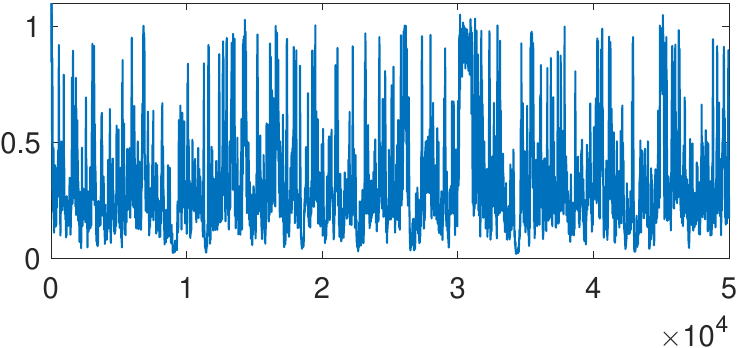}\vspace{-1.5\baselineskip}
	\caption{Traces for parameter $\lambda$ for the \textbf{pseudo-marginal slice sampler}. From top left to bottom right: $N=16$, $N=64$, $N=128$, $N=256$.}
	\label{fig:sinc2_slice}
\end{figure}

In Figures~\ref{fig:sinc2_pmhmc}--\ref{fig:sinc2_slice} below we show traces for the parameter $\lambda$ (which, along with $\sigma$, was the most difficult parameter to infer) for the various samplers considered with different settings.

%
%\begin{figure}[H]
%	\centering
%	\includegraphics[width=0.33\linewidth]{figs/hmc_scatter.png}%
%	\includegraphics[width=0.33\linewidth]{figs/pmhmc_scatter.png}%
%	\includegraphics[width=0.33\linewidth]{figs/pmslice_scatter.png}%
%	\caption{Scatter plots and marginal histograms for $(\sigma, \lambda)$ for
%		\textbf{standard HMC}, using a non-centered parameterisation (left),
%		\textbf{PM-HMC with $N=16$} (mid), and
%		\textbf{pseudo-marginal slice sampling with $N=128$} (right).}
%	\label{fig:scatter}
%\end{figure}

\subsection{Generalized linear mixed model}
In this section we give some additional details and results for the generalized linear mixed model considered in Section~\ref{sec:glmm} of the main manuscript.

The parameter values used for the data generation were:
\begin{align*}
	\beta = \begin{pmatrix}
	 -1.1671 &
	 2.4665 &
	 -0.1918 &
	 -1.0080 &
	 0.6212 &
	 0.6524 &
	 1.5410 &
	 0.2653
	\end{pmatrix}^T,
\end{align*}
$\mu_1=0$, $\mu_2 = 3$, $\lambda_1 = 10$, $\lambda_2 = 3$, and $w_1 = 0.8$.

All methods were initialized at the same point in $\theta$-space, as follows:
$\beta^{\text{init}}$ was sampled from $\N(0_p,I_p)$, resulting in:
\begin{align*}
	\beta^{\text{init}} = \begin{pmatrix}
	0.5838 &
	0.3805 &
	-1.5062 &
	-0.0442 &
	0.4717 &
	-0.1435 &
	0.6371 &
	-0.0522
	\end{pmatrix}^T,
\end{align*}
whereas the remaining parameters were initialized deterministically as
$\mu_1^{\text{init}} = 0$, $\mu_2^{\text{init}} = 0$
$\lambda_1^{\text{init}} = 1$, $\lambda_2^{\text{init}} = 0.1$,
and $w_1^{\text{init}} = 0.5$.

We used a $\N(0, 100)$ prior for each component of $\theta$.
However, for the particle Gibbs sampler we used a different parameterisation and (uninformative) conjugate priors when possible to ease the implementation. Varying the prior did not have any noticeable effect on the (poor) mixing of the Gibbs sampler.

For PM-HMC and the pseudo-marginal slice sampler we used a simple (indeed, naive) choice of importance distribution for the latent variables: $q(x_i) = \N(x_i \mid 0, 3^2)$ since this was easily represented in terms of Gaussian auxiliary variables (which is a requirement for both methods). A possibly better choice, which however we have not tried in practice, is to use a Gaussian or $t$-distributed approximation to the posterior distribution of the latent variables.
For the particle Gibbs sampler, which does not require the proposal to be represented in terms of Gaussian auxiliary variables, we instead used the (slightly better) proposal consisting of sampling from the prior for $X_i$.

The pseudo-marginal slice sampler made use of elliptical slice sampling
for updating the auxiliary variables, as recommended by \cite{MurrayG:2015}.
The components of $\theta$ were updated one-at-a-time using random walk Metropolis-Hastings kernels. We also updating $\theta$ jointly, but this resulted in very poor acceptance rates. The random-walk proposals were tuned to obtain acceptance rates of around 0.2--0.3.

The particle Gibbs sampler used conditional importance sampling kernels for the latent variables $X_{1:T}$, and for the parameters random walk Metropolis-Hastings kernels  were used when no conjugate priors were available.

\begin{figure}[H]
	\centering
	\includegraphics[width=0.5\linewidth]{mu_pmhmcN128}%
	\includegraphics[width=0.5\linewidth]{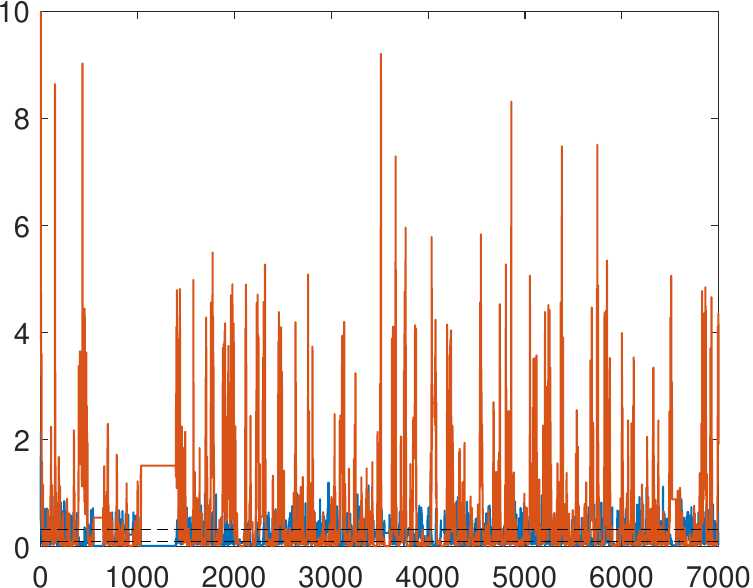}\vspace{-0.5\baselineskip}
	\caption{Traces for parameters $\mu_1$ and $\mu_2$ (left) and $1/\lambda_1$
		and $1/\lambda_2$ (right) for the \textbf{pseudo-marginal HMC sampler}
		with $N=128$.}
	\label{fig:glmm_pmhmc_trace}
\end{figure}

\begin{figure}[H]
	\centering
	\includegraphics[width=0.5\linewidth]{mu_sliceN128}%
	\includegraphics[width=0.5\linewidth]{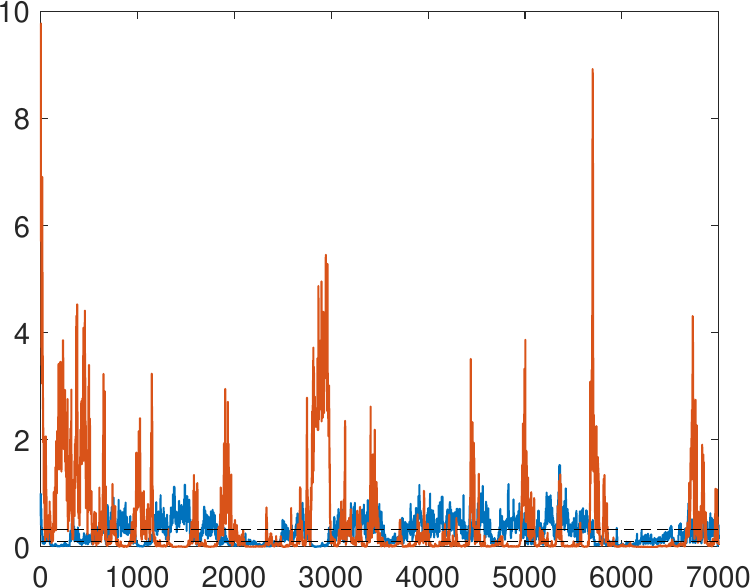}\vspace{-0.5\baselineskip}
	\caption{Traces for parameters $\mu_1$ and $\mu_2$ (left) and $1/\lambda_1$
		and $1/\lambda_2$ (right) for the \textbf{pseudo-marginal slice sampler}
		with $N=128$.}
	\label{fig:glmm_slice_trace}
\end{figure}

\begin{figure}[H]
	\centering
	\includegraphics[width=0.5\linewidth]{mu_gibbsN128}%
	\includegraphics[width=0.5\linewidth]{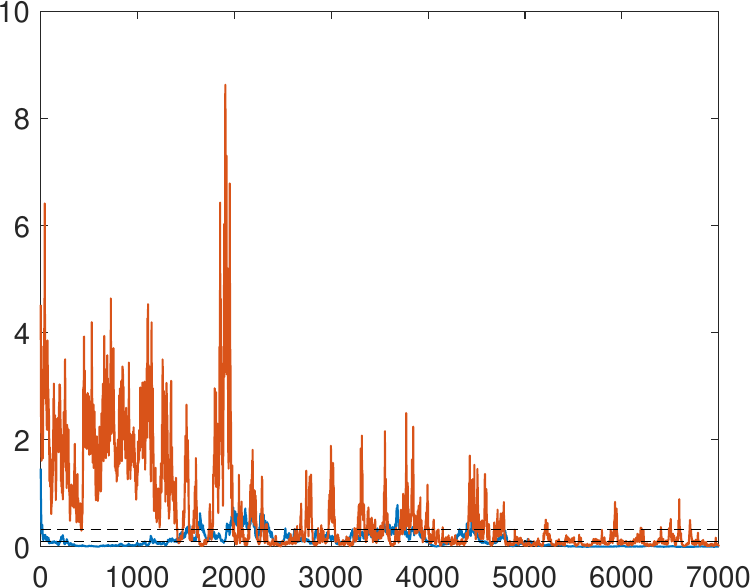}\vspace{-0.5\baselineskip}
	\caption{Traces for parameters $\mu_1$ and $\mu_2$ (left) and $1/\lambda_1$
		and $1/\lambda_2$ (right) for the \textbf{Gibbs sampler}
		using conditional importance sampling kernels with $N=128$ particles to update the latent variables.}
	\label{fig:glmm_gibbs_trace}
\end{figure}

\begin{figure}[H]
	\centering
	\includegraphics[width=0.33\linewidth]{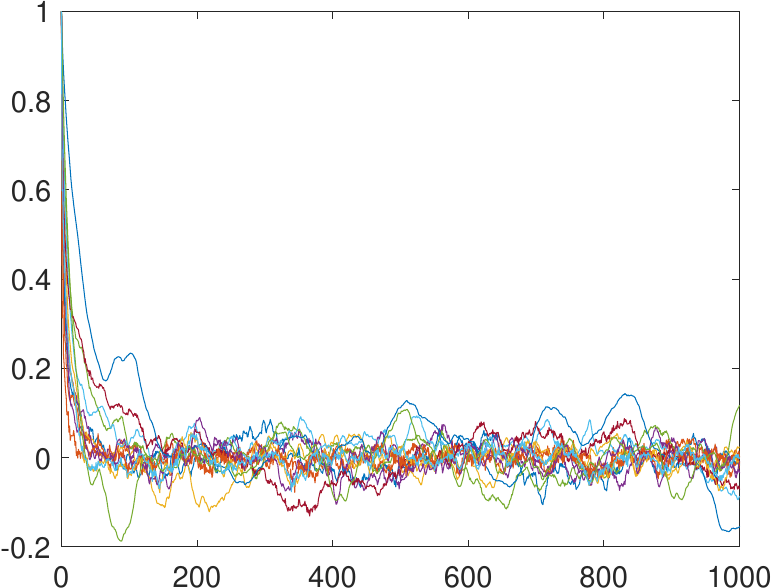}%
	\includegraphics[width=0.33\linewidth]{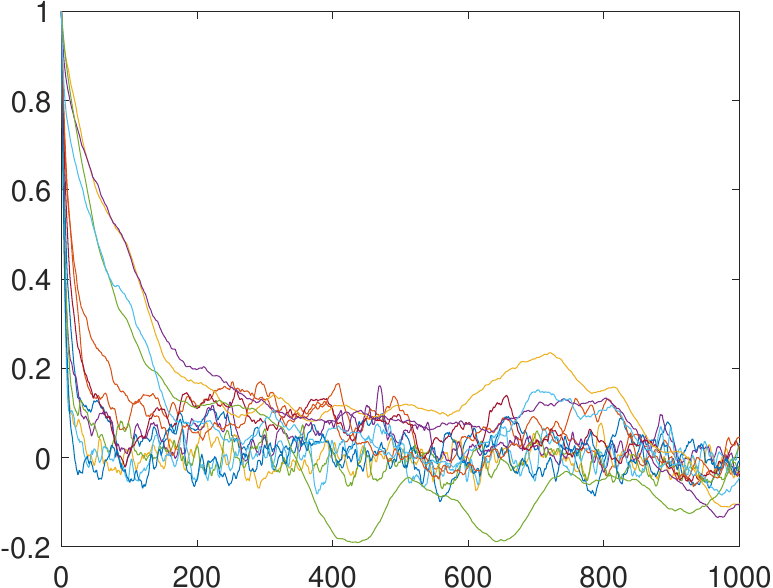}%
		\includegraphics[width=0.33\linewidth]{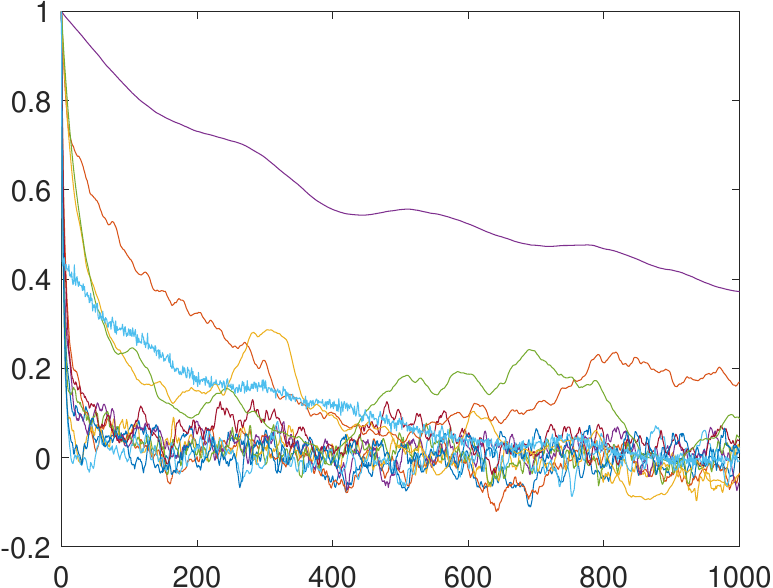}
	\caption{Estimated autocorrelations for the 13 parameters of the model for
		the \textbf{pseudo-marginal HMC sampler}
		with $N=128$ (left), \textbf{pseudo-marginal slice sampler} with $N=128$ (mid), and \textbf{Gibbs sampler} using conditional importance sampling kernels with $N=128$ particles to update the latent variables (right).}
	\label{fig:glmm_acf}
\end{figure}

Figures~\ref{fig:glmm_pmhmc_trace}--\ref{fig:glmm_gibbs_trace} show trace plots for the four parameters $\mu_1$, $\mu_2$, $1/\lambda_1$ and $1/\lambda_2$ of the Gaussian mixture model used to model the distribution of the random effects. The three plots correspond to the pseudo-marginal HMC sampler, the pseudo-marginal slice sampler, and the Gibbs sampler with conditional importance sampling kernels, respectively.
In Figure~\ref{fig:glmm_acf} we show estimated autocorrelations for the 13 parameters of the model for the three samplers.

%\begin{figure}[H]
%	\centering
%	\includegraphics[width=0.5\linewidth]{acf_sliceN128}
%	\caption{Estimated autocorrelations for the 13 parameters of the model for
%		the \textbf{pseudo-marginal slice sampler}
%		with $N=128$.}
%	\label{fig:glmm_slice_acf}
%\end{figure}
%\begin{figure}[H]
%	\centering
%	\includegraphics[width=0.33\linewidth]{acf_gibbsN128}
%	\caption{Estimated autocorrelations for the 13 parameters of the model for
%		the \textbf{Gibbs sampler}
%		using conditional importance sampling kernels with $N=128$ particles to update the latent variables.}
%	\label{fig:glmm_gibbs_acf}
%\end{figure}

\section{Auxiliary results}
For the proof of Proposition~\ref{prop:clt} we need to results to establish a CLT and LLN for dependent random variables. These results are given here to shorten the proofs above.
\begin{lemma}\label{lem:clt}
	Let $\{X_{i}\}_{i \in \nset}$ and $\{Y_i\}_{i \in \nset}$ be two sequences of random variables and $f$ a continuous function with continuous and bounded first derivative which satisfies $|f'|<c$. Let $h$ be a continuous bounded function with $|h|<d$ for some constant $d \in \rset^+$ and $g$ be a strictly positive continuous bounded function which satisfies $e^{-1} < g < e$ for some $e > 1$. Assume that there exists a random variable $Z$ such that
	\begin{align*}
		\frac{1}{\sqrt{N}} \sum_{i=1}^{N} f(X_i) \convd Z, \quad \text{as } N \to \infty.
	\end{align*}
	Then we have that
	\begin{align*}
		\frac{1}{\sqrt{N}} \sum_{i=1}^{N} f(X_i + \tfrac{h(Y_i)}{\sum_{j=1}^{N}g(Y_j)}) \convd Z, \quad \text{as }N \to \infty.
	\end{align*}

\end{lemma}

\begin{proof}
For any $\omega \in \Omega$ we have by Taylor's theorem that there exists a 
$$\bar{X}_i \in \left[ X_i - \tfrac{h(Y_i)}{\sum_{j=1}^{\N} g(Y_j)}, X_i + \tfrac{h(Y_i)}{\sum_{j=1}^{\N} g(Y_j)}\right]$$
 such that
	\begin{align*}
	  	f(X_i + \tfrac{h(Y_i)}{\sum_{j=1}^{N}g(Y_j)}) = f(X_i) + f'(\bar{X}_i) \tfrac{h(Y_i)}{\sum_{j=1}^{N}g(Y_j)}.
	  \end{align*}

	  Now we look at
	  \begin{align*}
	  	\frac{1}{\sqrt{N}}  \sum_{i=1}^{N} f(X_i + \tfrac{h(Y_i)}{\sum_{j=1}^{N}g(Y_j)}) = \frac{1}{\sqrt{N}} \sum_{i=1}^{N} f(X_i) + \frac{1}{\sqrt{N}} \sum_{i=1}^{N} f'(\bar{X}_i) \tfrac{h(Y_i)}{\sum_{j=1}^{N}g(Y_j)},
	  \end{align*}
	  where we use the results above to get the expression on the right hand side. We see that by the assumption the first term converges in distribution to $Z$ while the second term is something that we need to control. From the assumptions we have that
	  \begin{align*}
	  	- \frac{1}{\sqrt{N}} \sum_{i=1}^{N} \tfrac{c \cdot d}{\sum_{j=1}^{N} e^{-1}} < \frac{1}{\sqrt{N}} \sum_{i=1}^{N} f'(\bar{X}_i) \tfrac{h(Y_i)}{\sum_{j=1}^{N}g(Y_j)} < \frac{1}{\sqrt{N}} \sum_{i=1}^{N} \tfrac{c \cdot d}{\sum_{j=1}^{N} e^{-1}}.
	  \end{align*}
	  Since we have that
	  \begin{align*}
	  	\frac{1}{\sqrt{N}} \sum_{i=1}^{N} \tfrac{c \cdot d}{\sum_{j=1}^{N} e^{-1}} = \frac{1}{\sqrt{N}} c \cdot d \cdot e \to 0 \quad \text{as } N \to \infty,
	  \end{align*}
	  we have that, by the sandwich property
	  \begin{align*}
	  	\frac{1}{\sqrt{N}} \sum_{i=1}^{N} f'(\bar{X}_i) \tfrac{h(Y_i)}{\sum_{j=1}^{N}g(Y_j)} \convp 0, \quad \text{as } N \to \infty.
	  \end{align*}
	  The proof concludes by using Slutsky's lemma.\end{proof}

\begin{lemma}\label{lem:lln}
	Let $\{X_{i}\}_{i \in \nset}$ and $\{Y_i\}_{i \in \nset}$ be two sequences of random variables and $f$ a continuous function with continuous and bounded first derivative which satisfies $|f'|<c$. Let $h$ be a continuous bounded function with $|h|<d$ for some constant $d \in \rset^+$ and $g$ be a strictly positive continuous bounded function which satisfies $e^{-1} < g < e$ for some $e > 1$. Assume that there exists a constant $\mu$ such that
	\begin{align*}
		\frac{1}{{N}} \sum_{i=1}^{N} f(X_i) \convp \mu, \quad \text{as } N \to \infty.
	\end{align*}
	Then we have that
	\begin{align*}
		\frac{1}{{N}} \sum_{i=1}^{N} f(X_i + \tfrac{h(Y_i)}{\sum_{j=1}^{N}g(Y_j)}) \convp \mu, \quad \text{as }N \to \infty.
	\end{align*}
\end{lemma}

The proof of this is analogous with the previous proof and is therefore omitted.